\documentclass[aps,prb,twocolumn,showpacs,10pt,floatfix,longbibliography]{revtex4-2}

\usepackage{times}

\usepackage{graphicx}
\usepackage{subfigure}
\usepackage{amsmath}
\usepackage{amsfonts}

\usepackage{mathrsfs}
\usepackage{bbm}
\usepackage{subfigure}
\usepackage{xcolor}
\usepackage[colorlinks=true, urlcolor=blue, linkcolor=blue, citecolor=blue]{hyperref}

\usepackage[T1]{fontenc}

\usepackage{siunitx}
\usepackage{bm}
\usepackage[version=4]{mhchem}

\usepackage{stackengine}[2013-09-11]
\newcommand\slashzero{\stackinset{c}{}{c}{}{/}{0}}

\usepackage{dsfont}

\AddToHook{cmd/appendix/before}{%
  \setcounter{equation}{0}%
  \setcounter{theorem}{0}%
}

\begin{document}

\title{Coupled spin and valley Hall effects driven by coherent tunneling}

\author{W.~Zeng}
\email[E-mail: ]{zeng@ujs.edu.cn}
\affiliation{Department of physics, Jiangsu University, Zhenjiang 212013, China}

\begin{abstract}
We predict the coexistence of tunneling spin and valley Hall effects when electrons in graphene coherently transmit through a barrier with the broken inversion symmetry and proximity-induced spin-orbit coupling. Due to the rotation of the pseudospin in the tunneling process, the transmitted electrons acquire a finite spin- and valley-dependent backreflection geometric phase when the two interfaces of the barrier are asymmetric. This results in a spin- and valley-dependent skew coherent tunneling, which is responsible for the transverse spin and valley Hall currents. We further demonstrate that the coherent-tunneling assisted charge-spin and charge-valley conversions are highly efficient with large Hall angles. Our work provides a new route for the generation of efficient spin and valley Hall effects, suggesting potential applications for spintronic and valleytronic devices. 

\end{abstract}
\maketitle

\section{Introduction}\label{intro}

The possibility of using the spin and valley degrees of freedom to store and carry information leads to the emergence of spintronics and valleytronics \cite{RevModPhys.76.323,RevModPhys.90.021001,pulizzi2012spintronics}, respectively, where the spin Hall effect (SHE) and valley Hall effect (VHE) have attracted much attention in recent years. Both SHE and VHE can be generated through either extrinsic or intrinsic mechanisms. The extrinsic SHE is attributed to the spin-dependent scattering of charge carriers by impurities in the presence of the spin-orbit coupling \cite{PhysRevLett.112.066601,PhysRevLett.85.393}, whereas the intrinsic SHE is absent of any scattering processes and stems from the spin-orbit coupling in the electronic band structure \cite{PhysRevLett.92.126603,PhysRevLett.100.096401}. The extrinsic VHE can be realized by breaking the time-reversal symmetry, such as through the extrinsic magnetism caused by the magnetic proximity effect \cite{PhysRevB.92.121403} and the external light fields \cite{mak2014valley,lundt2019optical}. The intrinsic VHE is directly linked to the band topology, and the Hall conductance is proportional to the integration over the Fermi sea of the Berry curvature of each occupied band \cite{PhysRevLett.99.236809,PhysRevLett.75.1348}. The coupled SHE and VHE have been proposed in monolayers of $\ce{MoS}_2$ and  other group-VI dichalcogenides \cite{PhysRevLett.108.196802}, where the VHE, accompanied by a SHE, can be generated intrinsically by spin- and valley-dependent Berry curvature, or extrinsically by external optical fields. Recently, the valley-dependent SHE has been predicted in monolayer transition-metal dichalcogenides \cite{PhysRevLett.124.166803}, where the valley-polarized spin currents and transverse spin accumulations can be generated by the microwave irradiation.

Apart from the aforementioned methods, the skew tunneling in heterojunctions is an alternative mechanism to generate the SHE and VHE in the ballistic regime, which are termed as the \textit{tunneling} Hall effects. The tunneling spin Hall effect (TSHE) has been reported in magnetic tunnel junctions with broken time-reversal symmetry \cite{PhysRevLett.115.056602,PhysRevLett.110.247204,PhysRevLett.117.166806} and the tunneling valley Hall effect (TVHE) has been predicted in the tunnel junctions with tilted Dirac cones \cite{PhysRevLett.131.246301,PhysRevB.110.024511}. Both the existing TSHE and TVHE are associated with the momentum filtering and are independent of the Berry curvature.

In contrast to the previously predicted TSHE and TVHE, here we propose a tunneling Hall effect driven by the phase-coherent tunneling in graphene heterojunctions. Both the TSHE and TVHE can be generated without the need for breaking time-reversal symmetry or in the absence of tilted Dirac cones. Compared with the existing TSHE and TVHE, the physical origin of the predicted coherent-tunneling assisted Hall effect is attributed to the coherence of the tunneling acquired geometric phase rather than the time-reversal symmetry breaking induced momentum filtering \cite{PhysRevLett.115.056602,PhysRevLett.110.247204,PhysRevLett.117.166806} or the tilt-induced Fermi surface mismatch \cite{PhysRevLett.131.246301,PhysRevB.110.024511}. In addition, the previously predicted TSHE and TVHE are independent of the Berry curvature, whereas the non-$\pi$ Berry flux is essential in our model. We further demonstrate that the coherent-tunneling assisted charge-spin and charge-valley conversions are highly efficient with large Hall angles, suggesting potential applications for spintronic and valleytronic devices. 

The remainder of the paper is organized as follows. The model Hamiltonian and the scattering approach are explained in detail in Sec.\ \ref{scattering}. The backreflection phase is obtained in Sec.\ \ref{back}. The numerical results and discussions are presented in Sec.\ \ref{hall}. Finally, we conclude in Sec.\ \ref{conc}.

\section{Scattering approach}\label{scattering}

\begin{figure}[tp]
\begin{center}
\includegraphics[clip = true, width =\columnwidth]{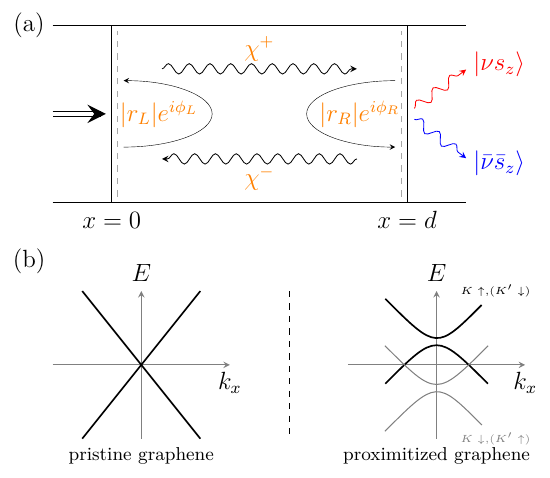}
\end{center}
\caption{(a) Schematic of the tunneling process. The tunnel junction is divided into three parts. The left and right electrode regions are located at $x<0$ and $x>d$, respectively. The barrier region is located at $0<x<d$, where the right and left propagating states are denoted by $\chi^+$ and $\chi^-$, respectively. The backreflections occur at the left and right interfaces of the barrier with the reflection amplitudes being $|r_{L}|e^{i\phi_L}$ and $|r_{R}|e^{i\phi_R}$, respectively. The transmitted state with valley and spin indices $|\nu,s_z\rangle$ and its time-reversal counterpart $|\bar{\nu}\bar{s}_z\rangle$ ($\bar{\nu}=-\nu$, $\bar{s}_z=-s_z$) are denoted by the red and blue wavy lines, respectively. (b) The energy dispersion at $\mu_b=0$ in the absence of the substrate (left) and in the presence of the substrate (right).}
\label{fig:junction}
\end{figure}

We consider the graphene-based tunnel junction in the $x-y$ plane, as shown in Fig.\ \ref{fig:junction}(a). The junction is divided into the left electrode region ($x<0$), right electrode region ($x>d$), and the central barrier region ($0<x<d$). In the left and right electrode regions, the electric transport property is described by the low-energy Hamiltonian \cite{RevModPhys.80.1337}
\begin{gather}
\mathcal{H}_{L/R}=v_F(\nu p_x\sigma_x+p_y\sigma_y)-\mu_{L/R},\label{eq:Hamiltonian}
\end{gather}
where $v_F$ is the Fermi velocity, $\nu=+$ ($-$) for $K$ ($K^\prime$) valley, $p_{x,y}=-i\partial_{x,y}$ are the two-dimensional momenta in the $x-y$ plane (we set $\hbar=1$), and $\sigma_{x,y}$ are the Pauli matrices in the sublattice space. $\mu_{L/R}$ is the Fermi energy in the left/right electrode regions, which can be adjusted by doping or by a gate voltage. Theoretical and experimental investigations demonstrate that the pseudospin staggered potential and the spin-orbit coupling can be induced in graphene by the proximity effect of substrates such as transition metal dichalcogenides \cite{PhysRevB.97.075434,PhysRevB.106.035137,PhysRevB.106.125417,wang2015strong}. In our model, the graphene in the barrier region is placed on the multilayer $\ce{WS2}$ substrate, where the induced spin-orbit interactions are dominated by the valley-Zeeman spin-orbit coupling taking the form of $\lambda\nu s_z\sigma_0$ due to the broken sublattice symmetry \cite{wang2015strong,PhysRevB.101.165436,Yang_2016,PhysRevB.97.075434}. Here $s_z=+1$ ($-1$) for the up (down) spin of the electrons. This spin-orbit coupling generates an effective valley-dependent Zeeman field, resulting in opposite spin splittings in different valleys, and thus is termed the valley-Zeeman spin-orbit coupling. Consequently, the Hamiltonian in the barrier region is given by~\cite{wang2015strong,PhysRevB.105.165427,PhysRevB.108.134511}
\begin{gather}
\mathcal{H}_b=v_F(\nu p_x\sigma_x+p_y\sigma_y)-\mu_b+h_I,\label{eq:Hamiltonian_b0}\\
h_I=\Delta\sigma_z+\nu s_z\lambda\sigma_0.\label{eq:Hamiltonian_b}
\end{gather}
Here $\mu_b$ is the Fermi energy in the barrier region. $h_I$ is the substrate interaction term consisting of the staggered potential ($\Delta\sigma_z$) and the spin-orbit coupling ($\nu s_z\lambda\sigma_0$), which can be induced by the proximity effect of substrates in the barrier region. $\sigma_z$ is the $z$-component of the Pauli matrix in the sublattice space. The substrate interaction term breaks the inversion symmetry by the staggered potential $\Delta\sigma_z$ but preserves the time-reversal symmetry. Consequently, the state $|\nu s_z\rangle$ and its time-reversal counterpart $|\bar{\nu}\bar{s}_z\rangle$ ($\bar{\nu}=-\nu$, $\bar{s}_z=-s_z$) are degenerate with the energy dispersion $E=\pm\sqrt{(v_F|\mathbf{p}|)^2+\Delta^2}+\nu s_z\lambda-\mu_b$, where $\pm$ for the conduction and valence bands, respectively. The energy dispersion features two linearly dispersing branches that touch at the Dirac point in the absence of a substrate; see Fig.\ \ref{fig:junction}(b) (left panel). However, the valley-spin dependent gap with a Zeeman splitting of $2\lambda$ appears in the presence of the proximity-induced staggered potential and the spin-orbit coupling, which has opposite sign in the $K$ and $K^\prime$ valleys and leads to an out-of-plane spin polarization with the opposite polarization in each valley; see Fig.\ \ref{fig:junction}(b) (right panel).

In order to obtain the net transmission probability across the barrier, we divide the whole tunneling process into two individual tunneling processes: (i) the tunneling through the left interface of the barrier and (ii) the tunneling through the right interface of the barrier, where the scattering amplitudes are $(t_{L},r_L)$ and $(t_R,r_R$), respectively. The net transmission amplitude across the barrier can be obtained by combining these two individual tunneling processes, which is given by the general Fabry-P\'erot form \cite{datta1997electronic,PhysRevLett.101.156804}
\begin{align}
t=\frac{t_Lt_R}{1-|r_Lr_R|e^{i\phi}}.\label{eq:fb}
\end{align}
The transmission amplitudes $t_j$, $r_{j}$ ($j=L,R$) and the total backreflected phase $\phi=\arg(r_L)+\arg(r_R)$ can be obtained by matching the wave functions at $x=0$ and $x=d$, resulting in the following equations:
\begin{gather}
    \chi^-+r_L\chi^+=t_L\tilde{\chi}_L^-,\label{eq:interface1}\\
    \chi^+e^{ip_bd}+r_R\chi^-e^{-ip_bd}=t_R\tilde{\chi}_R^+e^{ip_Rd}.\label{eq:interface2}
\end{gather}
Here $\chi^\pm=(\kappa\cos\varphi_b)^{-1/2}(\kappa e^{\mp i\nu\varphi_b/2},\pm e^{\pm i\nu\varphi_b/2})^T$ are the basis spinors in the barrier region with the superscript `$+$' and `$-$' denoting the right and left propagating mode, respectively; see Fig.\ \ref{fig:junction}(a). $\tilde{\chi}_{j}^+=(\cos\varphi_{j})^{-1/2}(e^{-i\nu\varphi_{j}/2},e^{i\nu\varphi_{j}/2})^T$ and $\tilde{\chi}_{j}^-=(\cos\varphi_{j})^{-1/2}(e^{i\nu\varphi_{j}/2},-e^{-i\nu\varphi_{j}/2})^T$ are the right and left propagating basis modes in the $j$ electrode ($j=L,R$), respectively. $p_{b}$ and $p_{R}$ are the longitudinal wave vectors in the barrier and right electrode regions, respectively. $\varphi_{j}$ ($j=L,R$) and $\varphi_b$ are the transmission angles in the $j$ electrode and the barrier, respectively, which are given by
\begin{align}
 \sin\varphi_j=v_Fp_y/(\epsilon+\mu_j),\quad \sin\varphi_b=v_Fp_y/\sqrt{\zeta^2-\Delta^2}\label{eq:angle}
\end{align}
with $p_y$ being the conserved transverse momentum labeling different trajectories and $\epsilon$ being the incident energy. The spin- and valley-dependent parameters $\kappa$ and $\zeta$ in Eq.\ (\ref{eq:angle}) are given by
\begin{align}
    \kappa^2=\frac{\zeta+\Delta}{\zeta-\Delta},\quad \zeta=\epsilon+\mu_b-\nu s_z\lambda,\label{eq:kapp}
\end{align}
respectively. The scattering amplitudes are obtained by solving Eqs.\ (\ref{eq:interface1}) and (\ref{eq:interface2}), which are given by
\begin{gather}
    t_j=\frac{2(\kappa\cos\varphi_b\cos\varphi_j)^{1/2}}{e^{i\nu(-1)^{\ell_j+1}(\varphi_b+\varphi_j)/2}+\kappa e^{i\nu(-1)^{\ell_j}(\varphi_b+\varphi_j)/2}},\label{eq:transmission}\\
    r_j=\frac{e^{i(-1)^{\ell_j}\nu\varphi_b}-\kappa e^{i(-1)^{\ell_j}\nu\varphi_j}}{1+\kappa e^{i(-1)^{\ell_j}\nu(\varphi_b+\varphi_j)}}.\label{eq:reflection}
\end{gather}
Here $\nu=\pm1$ is the valley index, $\ell_j=0$ for $j=R$, and $\ell_j=1$ for $j=L$. One can verify that the identity $|t_j|^2+|r_j|^2=1$ holds due the conservation of the probability current. The tunneling probability for the $j$ interface can be obtained by squaring $t_j$ in Eq.\ (\ref{eq:transmission}), which gives rise to
\begin{gather}
    T_j=\frac{4\kappa\cos\varphi_b\cos\varphi_j}{1+\kappa^2+2\kappa\cos(\varphi_b+\varphi_j)}.\label{eq:probability}
\end{gather}
The reflection probability is given by $R_j=1-T_j$. It is noted that $T_j$ remains unchanged under either the substitution $p_y\rightarrow-p_y$ or $(\nu,s_z)\rightarrow(\bar\nu,\bar{s}_z)$ ($\bar{\nu}=-\nu$ and $\bar{s}_z=-s_z$). Consequently, the electrons with opposite valley and spin indices (\textit{i.e.}, $|\nu s_z\rangle$ and $|\bar{\nu} \bar{s}_z\rangle$) exhibit the same transport behaviors through the single interface tunneling process and the transmission probability is symmetric with respect to $p_y$. The transmission probability is independent of the backreflection phase, and the skew tunneling is absent during this single interface tunneling process. However, when electrons transmit through two combined interfaces, \textit{i.e.}, a barrier region, the coherence of the backreflection phases acquired at each interface of the barrier plays a key role in the transmission process, leading to a phase-coherent tunneling, which is explained in detail in Sec.\ \ref{hall}.

\section{Backreflection phase}\label{back}
\begin{widetext}
The total backreflection phase acquired at the barrier interfaces can be directly obtained by $\phi=\sum_{j=L,R}\arg(r_j)$. With the help of Eq.\ (\ref{eq:reflection}), we obtain
\begin{align}
    \arg(r_j)=\pi+2(1-\ell_j)p_pd-(-1)^{\ell_j}\arctan\Big[\frac{\kappa^2+1}{\kappa^2-1}\tan\varphi_b\Big]+\phi_{G,j},\label{eq:arg}
\end{align}
where
\begin{align}
    \phi_{G,j}=(-1)^{\ell_j}\arctan\Big[\frac{2\nu(\kappa^2-1)\big((\kappa^2+1)\sin\varphi_b-\kappa\sin\varphi_j\big)\cos\varphi_b}{\big((\kappa^4+1)\cos2\varphi_b-2\kappa^2\big)\sin\varphi_b+2\kappa(\kappa^2+1)\sin\varphi_b\sin\varphi_j}\Big].
\end{align}
The summation of the first term $\pi$ in Eq.\ (\ref{eq:arg}) at $L$ and $R$ interfaces results in a $2\pi$ phase shift, which makes no contributions to the coherent tunneling and can be omitted here for simplicity. Both the second and third terms in Eq.\ (\ref{eq:arg}) are gauge dependent, which can be removed by a gauge transformation for a single interface. However, for two interfaces with a fixed spacing $d$ (\textit{i.e.}, the barrier length is $d$), the summation of these terms gives rise to a kinetic phase due to the motion of electrons in the barrier:
\begin{align}
\phi_0=\sum_{j=L,R}\Big(2(1-\ell_j)p_pd+(-1)^{\ell_j}\arctan\Big[\frac{\kappa^2+1}{\kappa^2-1}\tan\varphi_b\Big]\Big)=2p_bd,\label{eq:phi0}
\end{align}
where $p_b=(\zeta^2-\Delta^2-v_F^2p_y^2)^{-1/2}$ is the longitudinal momentum in the barrier region. The last term $\phi_{G,j}$ ($j=L,R$) in Eq.\ (\ref{eq:arg}) is the additional phase acquired at $j$ interface, which disappears when the inversion symmetry is preserved ($\kappa=1$). The total additional phase shift, expressed as $\phi_G=\sum_{j=L,R}\phi_{G,j}$, is given by 
\begin{align}
    \tan\phi_G=&\nu\times\frac{2\kappa(\kappa^2-1)(\sin\varphi_L-\sin\varphi_R)\cos\varphi_b}{1+\kappa^4-2\kappa^2\cos2\varphi_b+4\kappa^2\sin\varphi_L\sin\varphi_R-2\kappa(1+\kappa^2)(\sin\varphi_L+\sin\varphi_R)\sin\varphi_b}.\label{eq:geo}
\end{align}
Abbreviating $c_b=v_F/(\zeta^2-\Delta^2)^{1/2}$, $c_{L/R}=v_F/(\epsilon+\mu_{L/R})$, and with the help of Eqs.\ (\ref{eq:angle})-(\ref{eq:kapp}), Eq.\ (\ref{eq:geo}) can be rewritten in a compact form
\begin{align}
\tan\phi_{G}=&f_{\nu s_z}\times \nu\Delta  (\mu_R-\mu_L)p_y,\label{eq:pG}
\end{align}
where 
\begin{align}
f_{\nu s_z}=\frac{4\kappa v_Fp_b c_bc_Lc_R(\zeta-\Delta)^{-1}}{1+\kappa^4+4\kappa^2c_Lc_Rp_y^2+2\kappa^2c_b^2(p_y^2-p_b^2)-2c_b\kappa(1+\kappa^2)(c_L+c_R)p_y^2}.\label{eq:fff}
\end{align}
\end{widetext}
Here $f_{\nu s_z}$ is an even function of $p_y$ and satisfies $f_{\nu s_z}=f_{\bar{\nu} \bar{s}_z}$. Consequently, the total phase shift of the transmission amplitude in Eq.\ (\ref{eq:fb}) is given by 
\begin{align}
    \phi=\sum_{j=L,R}\arg(r_j)=\phi_0+\phi_G.
\end{align}

It is noted that $\phi_G$ is dependent on the junction parameters and disappears when either $\Delta=0$ or $\mu_L=\mu_R$; see Eq.\ (\ref{eq:pG}). In addition, distinct from the kinetic phase $\phi_0$, $\phi_G$ changes sign under the substitution $p_y\rightarrow-p_y$, namely
\begin{align}
   p_y\rightarrow-p_y\implies
    \left\{
        \begin{array}{l}
         \phi_0\rightarrow\phi_0, \\
          \phi_G\rightarrow-\phi_G.
        \end{array}
    \right.\label{eq:imply}
\end{align}

\begin{figure}[tp]
\begin{center}
\includegraphics[clip = true, width =0.85\columnwidth]{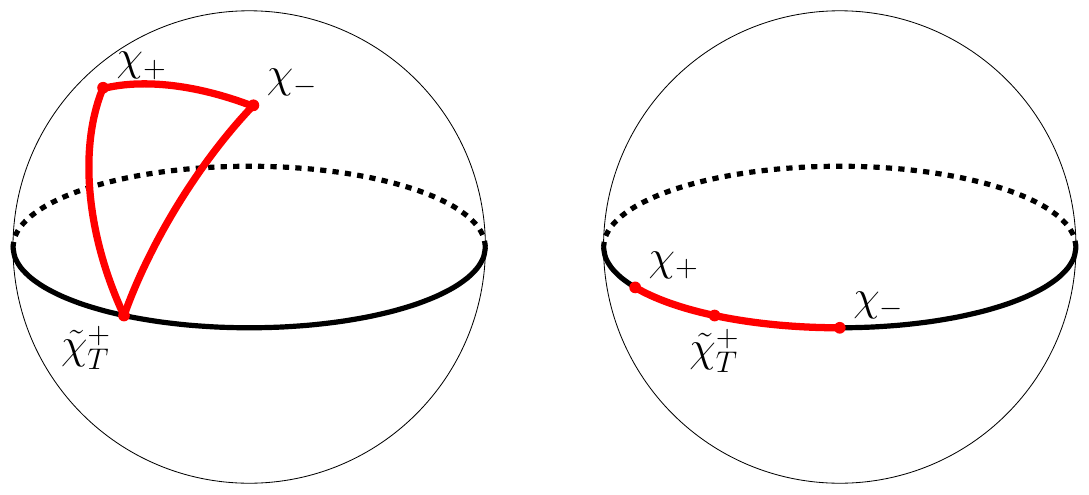}
\end{center}
\caption{Polarization vectors on the Bloch sphere. (Left) $\Phi\neq\pi$. (Right) $\Phi=\pi$.  }
\label{fig:ball}
\end{figure}

The additional phase shift $\phi_G$ originates from the rotation of the pseudospin of the Dirac fermions in the reflection at the left and right interfaces of the barrier, which is termed as the geometric phase \cite{RevModPhys.64.51,RevModPhys.66.899,anandan1992geometric}. Contrary to the usual Berry phase, the backreflection geometric phase can be acquired by the single electron tunneling process through the interfaces of the barrier, where the incident, reflected, and transmitted states can be mapped onto the Bloch sphere by the pseudospin polarization vectors. This acquired geometric phase can be electrically tuned by the junction control, leading to the geometric-phase devices such as the topological waveguides and topological transistor \cite{PhysRevB.87.165420}.

We follow the works in Refs. \cite{PhysRevB.87.165420,PhysRevA.93.042124} to provide a detailed explanation of the geometrical origin of $\phi_G$. The backreflection phase can be directly calculated from Eqs.\ (\ref{eq:interface1}) and (\ref{eq:interface2}). Taking the right interface as an example, 
by applying $\tilde{\chi}^+_{T}$ to both side of Eq.\ (\ref{eq:interface2}), where $\tilde{\chi}^+_{T}$ is the spinor orthogonal to $\tilde{\chi}^+$, one finds 
\begin{align}
    r_R=-e^{2ip_bd}\frac{\langle\tilde{\chi}^+_{T}|\chi^+\rangle}{\langle\tilde{\chi}^+_{T}|\chi^-\rangle}.\label{eq:rraa}
\end{align}
With the help of the geodesic rule of $\arg\langle\chi_a|\chi_b\rangle=i\int_\mathcal{C}d\mathbf{s}\cdot\langle\chi_\mathbf{s}|\bm\nabla|\chi_\mathbf{s}\rangle$ (the integration path $\mathcal{C}$ is along the geometric line from $\chi_b$ to $\chi_a$ on the Bloch sphere), the phase of $r_R$ is given by \cite{PhysRevB.89.155412}
\begin{align}
    \arg(r_R)=\pi+2p_bd-\arg\langle\chi^+|\chi^-\rangle-\frac{\Omega_{R}}{2},
\end{align}
where $\pi$ corresponds to the first term of the backreflection phase derived in Eq.\ (\ref{eq:arg}), $2p_bd$ and $-\arg\langle\chi^+|\chi^-\rangle$ are gauge dependent and correspond to the second and third terms in Eq.\ (\ref{eq:arg}), respectively. The last term $-\Omega_{R}/2$ corresponds to $\phi_{G,R}$ in Eq.\ (\ref{eq:arg}) with $\Omega_{R}$ being the solid angle covered by the geodesic triangle connecting $\chi^+$, $\tilde{\chi}^+_{T}$, and $\chi^-$ on the Bloch sphere, which gives rise to $\phi_{G,R}=-\Omega_{R}/2$. 

The given spinor $\chi$ can be mapped onto the Bloch sphere by the pseudospin polarization vector, which is given by $\mathbf{d}(\chi)=\langle\chi|\bm\sigma|\chi\rangle$ with $\bm\sigma$ being the spin-$1/2$ Pauli matrices. In the right electrode region, the polarization vector of $\tilde{\chi}^+_{T}$ is given by 
\begin{align}
    \mathbf{d}_R=(\cos\varphi_R,-\sin\varphi_R,0)^T,
\end{align}
which always lies on the equator of Bloch sphere. In the barrier, the polarization vectors of $\chi^+$ and $\chi^-$ are given by 
\begin{gather}
    \mathbf{d}_\pm=\Big(\pm\frac{2\kappa\cos\varphi_b}{\kappa^2+1},\frac{2\kappa\sin\varphi_b}{\kappa^2+1},\frac{\kappa^2-1}{\kappa^2+1}\Big)^T,
\end{gather}
respectively. A $\kappa$-dependent $z$ component of $\mathbf{d}_\pm$ appears, which can be written as $d_z=(\kappa^2-1)/(\kappa^2+1)=\Phi/\pi-1$ with $\Phi=2\pi\kappa^2/(1+\kappa^2)$ being the spin- and valley-dependent Berry flux \cite{RevModPhys.82.1959,PhysRevB.84.205440}. The nonzero $d_z$ appears when the inversion symmetry is broken in the barrier region ($\Phi\neq\pi$), leading to a finite solid angle covered by the geodesic triangle connecting $\chi^+$, $\chi^-$, and $\tilde{\chi}^+_{T}$ on the Bloch sphere; see Fig.\ \ref{fig:ball} (left panel). Consequently, a finite geometric phase is acquired in this situation. However, when the inversion symmetry is preserved ($\Phi=\pi$), $\chi^+$, $\chi^-$, and $\tilde{\chi}^+_{T}$ all lie on the equator of the Bloch sphere, as shown in Fig.\ \ref{fig:ball} (right panel), where the solid angle vanishes, leading to the absence of the geometric phase. For the left interface, the acquired solid angle $\Omega_L$ can be obtained by the same method, giving rise to the additional geometric phase at the left interface being $\phi_{G,L}=-\Omega_{L}/2$. Consequently, the total geometric phase is given by $\phi_G=\phi_{G,L}+\phi_{G,R}=-(\Omega_L+\Omega_R)/2$. The pseudospin polarization of the state orthogonal to $\tilde\chi^{-}$ is given by $\mathbf{d}_L=(\cos\varphi_L,\sin\varphi_L,0)^T$, which also always lies on the equator of Bloch sphere. As a result, $\Omega_L$ is also absent when the inversion symmetry is preserved. Moreover, the symmetric interfaces of the barrier (\textit{i.e.}, $\mu_L=\mu_R$) result in $\varphi_L=\varphi_R$ and thus $\Omega_L=-\Omega_R$, where the total geometric phase is also absent in this situation. 

Consequently, a nonzero total geometric phase $\phi_G$ requires two essential conditions: (i) inversion symmetry breaking ($\Delta\neq0$), ensuring that both $\Omega_L$ and $\Omega_R$ are nonzero, and (ii) asymmetric barrier interfaces ($\mu_L\neq\mu_R$), ensuring that $\Omega_L\neq-\Omega_R$, which are in agreement with the analytical expression of $\phi_G$ in Eq.\ (\ref{eq:pG}).

We note that the backreflection geometric phase is acquired at the interfaces between the barrier and the electrodes in our model. This is different from the geometric phase acquired at the edge of the graphene sheet, where the results are dependent on the edge geometries (armchair or zigzag) due to the fact that different edge geometries possess different boundary conditions \cite{PhysRevB.89.155412}. However, since the staggered potential and the spin-orbit coupling in the barrier region are externally proximity-induced from the substrate, there is no lattice mismatch between the electrodes and the barrier. The boundary conditions at the interfaces of the barrier in our model require the continuity of the wave functions at the interfaces [Eqs.\ (\ref{eq:interface1}) and (\ref{eq:interface2})], corresponding to the conservation of the probability current traveling perpendicular to the barrier, which are independent of the interface geometries.

\begin{figure}[tp]
\begin{center}
\includegraphics[clip = true, width =\columnwidth]{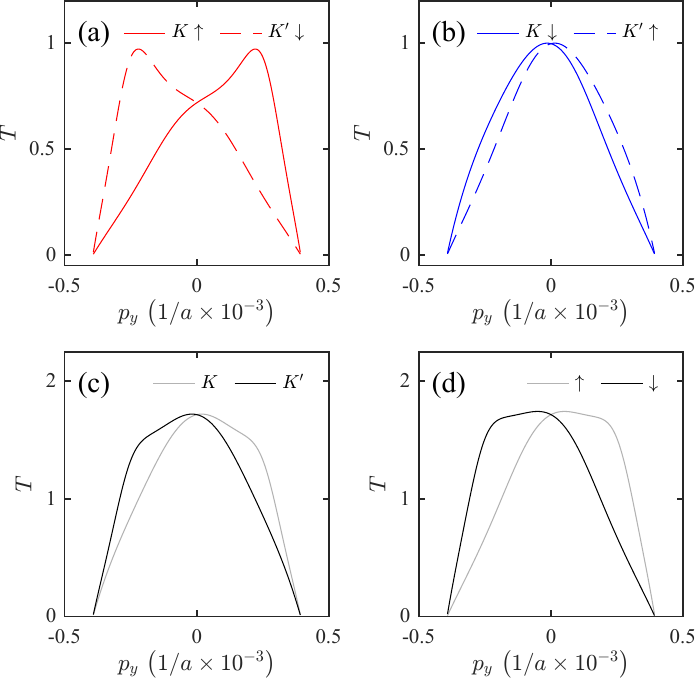}
\end{center}
\caption{Spin- and valley-resolved transmission probability versus the transverse momentum $p_y$ (in units of $1/a\times10^{-3}$ with $a$ being the lattice constant of graphene). The parameters are $\epsilon=\SI{1.8}{\meV}$, $\Delta=\SI{3.25}{\meV}$, $\lambda=\SI{2.5}{\meV}$, $\mu_L=\SI{1}{\meV}$, $\mu_R=\SI{-5.6}{\meV}$, $\mu_b=\SI{6.5}{\meV}$ and $d=\SI{25}{\nm}$. (a) Transmission probability for the state $|K\uparrow\rangle$ and its time-reversal counterpart $|K'\downarrow\rangle$. (b) Transmission probability for the state $|K\downarrow\rangle$ and its time-reversal counterpart $|K'\uparrow\rangle$. (c) Transmission probability for electrons from different valleys. (d) Transmission probability for electrons with different spins.}
\label{fig:trans}
\end{figure}

\section{Tunneling spin and valley Hall effects}\label{hall}

The net transmission probability through the barrier can be obtained by squaring $t$ in Eq.\ (\ref{eq:fb}), which is given by
\begin{align}
    T(p_y)=&\frac{T_LT_R}{1+R_LR_R-2\sqrt{R_LR_R}\cos(\phi_0+\phi_G)}\nonumber\\=&T_LT_R\sum_{n=0}^\infty \frac{(4R_LR_R)^{n/2}}{(1+R_LR_R)^{n+1}}\cos^n(\phi_0+\phi_G),\label{eq:tot}
\end{align}
where $T_{j}$ ($j=L,R$) is the individual transmission probability for the $j$ interface derived in Eq.\ (\ref{eq:probability}) and $R_j=1-T_j$ is the corresponding reflection probability. Abbreviating $g_n=T_LT_R(4R_LR_R)^{n/2}/(1+R_LR_R)^{n+1}$, and with the help of Eq.\ (\ref{eq:imply}), we obtain
\begin{gather}
    T(p_y)=\sum_{n} g_{n} \cos^n(\phi_0+\phi_{G}),\label{eq:t1}\\
    T(-p_y)=\sum_{n} g_{n} \cos^n(\phi_0-\phi_{G}),\label{eq:t2}
\end{gather}
which indicate that the nonzero $\phi_G$ yields $T(p_y)\neq T(-p_y)$, leading to a skew coherent tunneling. We note that the net transmission probabilities in Eqs.\ (\ref{eq:t1}) and (\ref{eq:t2}) are spin- and valley-dependent, namely $T(p_y)=T_{\nu s_z}(p_y)$, where the valley and spin indices `$\nu,s_z$' are omitted here for simplicity.

The transmission probability as a function of the transverse momentum $p_y$ is shown in Fig.\ \ref{fig:trans}. In our calculations, we take $\Delta=\SI{3.25}{\meV}$ and $\lambda=\SI{2.5}{\meV}$, which are consistent with the \textit{ab initio} calculations for the staggered potential and the spin-orbit coupling in graphene induced by proximity to the transition metal dichalcogenide substrates, such as $\ce{WS2}$ \cite{wang2015strong}. For the given valley and spin indices, $\phi_G$ changes sign when $p_y$ is reversed, leading to a large anisotropy for the $p_y$-resolved transmission probability; see Figs.\ \ref{fig:trans}(a) and \ref{fig:trans}(b). The transverse valley and spin imbalances occur due to $T_{\nu s_z}(p_y)\neq T_{\bar\nu s_z}(p_y)$ and $T_{\nu s_z}(p_y)\neq T_{\nu \bar{s}_z}(p_y)$, respectively, where $\bar\nu=-\nu$ and $\bar{s}_z=-s_z$. The net valley- and spin-dependent transmission probabilities (given by $T_{\nu}=\sum_{s_z}T_{\nu s_z}$ and $T_{s_z}=\sum_{\nu}T_{\nu s_z}$, respectively) are shown in Figs.\ \ref{fig:trans}(c) and \ref{fig:trans}(d). The valley-dependent transmission probability in Fig.\ \ref{fig:trans}(c) indicates that the electrons in $K$ valley have a large transmission for positive $p_y$, whereas the transmissions in $K^\prime$ valley are similarly asymmetric but skewed into the opposite direction, leading to the electrons with opposite valleys turning into different transverse directions, which is responsible for the net transverse valley currents. Similarly, the net transverse spin currents also occur due to the skew behaviors of $T_{s_z}$, as shown in Fig.\ \ref{fig:trans}(d). $\phi_G$ is even under time-reversal symmetry $(p_y,\nu,s_z)\rightarrow(-p_y,\bar\nu,\bar{s}_z)$, resulting in $T_{\nu s_z}(p_y)=T_{\bar\nu \bar{s}_z}(-p_y)$. Thus, the tunneling between $|\nu s_z\rangle$ and $|\bar\nu \bar{s}_z\rangle$ are always symmetric, as shown in Figs.\ \ref{fig:trans}(a) and \ref{fig:trans}(b), and the transverse charge currents are always absent.

\begin{figure}[tp]
\begin{center}
\includegraphics[clip = true, width =\columnwidth]{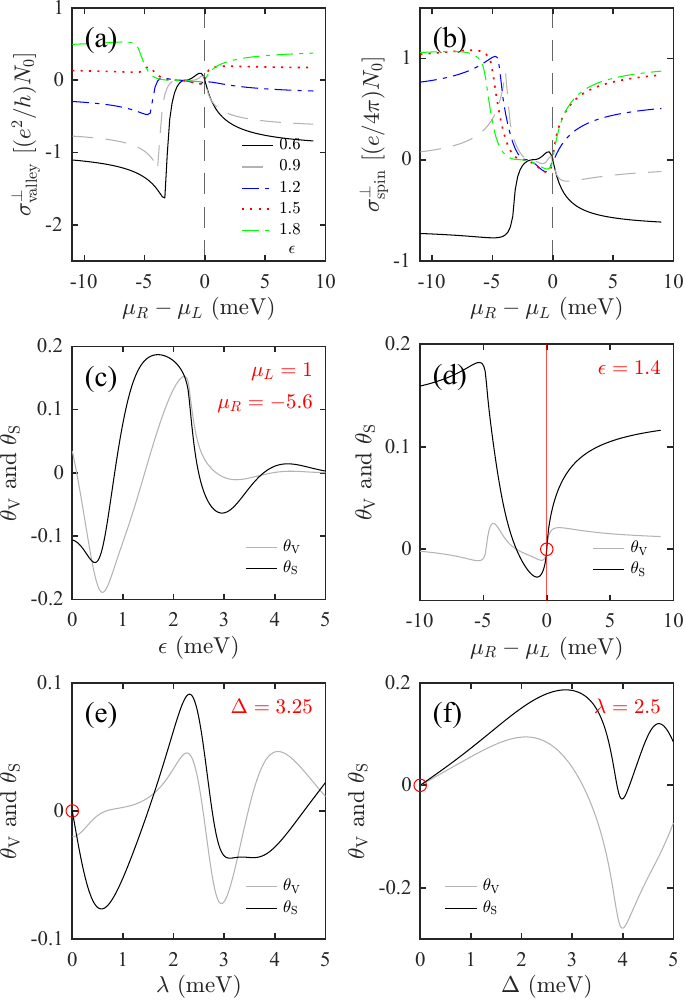}
\end{center}
\caption{[(a), (b)] The transverse valley and spin Hall conductance versus $\mu_R-\mu_L$ at different incident energy $\epsilon$ (in units of $\mathrm{meV}$). The other parameters in (a) and (b) are $\Delta=\SI{3.25}{\meV}$, $\lambda=\SI{2.5}{\meV}$, $\mu_b=\SI{6.5}{\meV}$, and $d=\SI{25}{\nm}$. The number of the transverse modes is $N_0=(\epsilon+\mu_L)W/\pi\hbar v_F$ with $W$ being the width of the junction. (c) Spin Hall angle $\theta_{\mathrm{S}}$ and valley Hall angle $\theta_{\mathrm{V}}$ versus the incident energy $\epsilon$ for $\mu_L=\SI{1}{\meV}$, $\mu_R=\SI{-5.6}{\meV}$. (d) Spin Hall angle $\theta_{\mathrm{S}}$ and valley Hall angle $\theta_{\mathrm{V}}$ versus $\mu_R-\mu_L$ for $\epsilon=\SI{1.4}{\meV}$. The red vertical line denotes $\mu_R-\mu_L=0$ and the red circle denotes $\theta_{\mathrm{S}}=\theta_{\mathrm{V}}=0$. The other parameters in (c) and (d) are $\Delta=\SI{3.25}{\meV}$, $\lambda=\SI{2.5}{\meV}$, $\mu_b=\SI{6.5}{\meV}$ and $d=\SI{25}{\nm}$. (e) Spin Hall angle $\theta_{\mathrm{S}}$ and valley Hall angle $\theta_{\mathrm{V}}$ versus $\lambda$ for $\Delta=\SI{3.25}{\meV}$. The red circle denotes the absence of TSHE ($\theta_{\mathrm{S}}=0$) at $\lambda=0$. (f) Spin Hall angle $\theta_{\mathrm{S}}$ and valley Hall angle $\theta_{\mathrm{V}}$ versus $\Delta$ for $\lambda=\SI{2.5}{\meV}$. The red circle denotes the absence of both TSHE and TVHE ($\theta_{\mathrm{S}}=\theta_{\mathrm{V}}=0$) at $\Delta=0$. The other parameters in (e) and (f) are $\mu_L=\SI{1}{\meV}$, $\mu_R=\SI{-5.6}{\meV}$, $\mu_b=\SI{6.5}{\meV}$ and $d=\SI{25}{\nm}$.}
\label{fig:theta}
\end{figure}

The transverse valley conductance [in units of $(e^2/h)$] and the transverse spin conductance [in units of $(e/4\pi)$] can be obtained within the Landauer formalism \cite{datta1997electronic,PhysRevLett.115.056602,PhysRevLett.117.166806,PhysRevB.110.024511}, which are given by
\begin{align}
     \sigma_\mathrm{valley}^\perp=\sum_{\nu s_z}\sum_{p_y}\nu\gamma T_{\nu s_z},\quad \sigma_\mathrm{spin}^\perp=\sum_{\nu s_z}\sum_{p_y}s_z\gamma T_{\nu s_z},
 \end{align} 
respectively. Here $\gamma=v_y/v_x$ is the ratio of the transverse group velocity to the longitudinal group velocity. We present the transverse conductance versus the difference of the right and left chemical potentials in Figs.\ \ref{fig:theta}(a) and \ref{fig:theta}(b). It is shown that both the transverse valley and spin conductance are sensitive to the incident energy and disappear at $\mu_R-\mu_L=0$ due to the absence of the total geometric phase in the barrier region. The efficiency of the charge-valley and charge-spin conversions can be characterized by the valley Hall angle ($\theta_{\mathrm{V}}$) and spin Hall angle ($\theta_{\mathrm{S}}$), respectively, which are given by \cite{PhysRevLett.115.056602,PhysRevLett.131.246301} $\tan\theta_{\mathrm{V}}= \sigma_{\mathrm{valley}}^\perp/\sigma^\parallel$ and $\tan\theta_{\mathrm{S}}= \sigma_{\mathrm{spin}}^\perp/\sigma^\parallel$ with $\sigma^\parallel=\sum_{\nu s_z}\sum_{p_y}T_{\nu s_z}$ being the longitudinal conductance [in units of $(e^2/h)$]. $\theta_{\mathrm{S}}$ and $\theta_{\mathrm{V}}$ versus the incident energy $\epsilon$ are shown in Fig.\ \ref{fig:theta}(c). The valley Hall angle approaches the maximum value at $\epsilon\simeq\qty{0.6}{\meV}$ with the absolute value of the valley Hall angle being $|\theta_{\mathrm{V}}|=0.2$. The value obtained here are comparable to that found in the tunneling valley Hall effect caused by the tilt mechanism, where $\theta_{\mathrm{V}}=0.15$ for a weak tilting Dirac cone \cite{PhysRevLett.131.246301}. However, the typical value of the valley Hall angle in our model is smaller than that in the pure crossed Andreev reflection assisted tunneling valley Hall effect, where the valley Hall angle can reach an order of unity \cite{PhysRevB.106.094503}. For the spin Hall effect, the spin Hall angles are usually very small in semiconductors (in the range of $0.0001$ to $0.001$) \cite{doi:10.1126/science.1105514,ando2012observation} but can be considerably enhanced by the single impurity scattering in graphene \cite{PhysRevLett.112.066601}, where the spin Hall angle is in the range of $0.01$ to $0.1$. However, a large spin Hall angle can be generated in our model, as shown in Fig.\ \ref{fig:theta}(c). $\theta_{\mathrm{S}}$ approaches the maximum value at $\epsilon\simeq\qty{1.5}{\meV}$ with the maximum value being $\theta_{\mathrm{S}}=0.2$, which is comparable to the large spin Hall angle obtained in graphene grown by chemical vapour deposition \cite{balakrishnan2014giant}. $\theta_{\mathrm{S}}$ and $\theta_{\mathrm{V}}$ versus the difference between the left and right chemical potentials $\mu_R-\mu_L$ are shown in Fig.\ \ref{fig:theta}(d). When the chemical potentials on both electrodes are equal, \textit{i.e.}, $\mu_R-\mu_L=0$, as denoted by the red vertical line in Fig.\ \ref{fig:theta}(d), both the TSHE and TVHE are absent due to the absence of the total geometric phase, giving rise to $\theta_{\mathrm{S}}=\theta_{\mathrm{V}}=0$, as indicated by the red circle marked in Fig.\ \ref{fig:theta}(d).

$\theta_{\mathrm{S}}$ and $\theta_{\mathrm{V}}$ versus the proximity-induced parameters are shown in Figs.\ \ref{fig:theta}(e) and \ref{fig:theta}(f). The pure TVHE can be generated when the spin-orbit coupling is absent, where a finite $\theta_{\mathrm{V}}$ appears at $\lambda=0$ with zero $\theta_{\mathrm{S}}$, as denoted by the red circle in Fig.\ \ref{fig:theta}(e). This pure TVHE is attributed to the valley-contrasting geometric phase. For $\lambda=0$, the spin-dependence of the parameter $f_{\nu s_z}$ is removed in Eq.\ (\ref{eq:pG}). Swapping the valley indices results in the sign change of $\phi_G$, which is responsible for the valley-contrasting skew tunneling. Both the TSHE and TVHE are absent when the inversion symmetry is preserved in the barrier ($\Delta=0$) due to the absence of the geometric phase, as denoted by the red circle in Fig.\ \ref{fig:theta}(f).

The tunneling Hall effects are always associated with the momentum filtering. Compared with the existing TSHE and TVHE, the origin of the momentum filtering in our model is different. The momentum filtering in the previously proposed TSHE is attributed to the time-reversal symmetry mismatch in the tunnel junction \cite{PhysRevLett.115.056602,PhysRevLett.110.247204,PhysRevLett.117.166806}, where the time-reversal symmetry is broken in magnetic electrodes but preserved in the nonmagnetic barrier layer with spin-orbit coupling. For the TVHE predicted in tilted Dirac and Weyl systems \cite{PhysRevLett.131.246301,PhysRevB.110.024511}, the momentum filtering occurs due to the mismatch of the Fermi surfaces of the tilted Dirac and Weyl fermions. However, the momentum filtering in our model occurs due to the transverse-momentum-dependent geometric phase shift, which is responsible for the skew coherent tunneling. In addition, distinct from the previously proposed Berry-curvature-free TSHE and TVHE, the non-$\pi$ Berry flux in the barrier region is essential in our model, which is introduced by breaking the inversion symmetry via the proximity-induced pseudospin staggered potential $\Delta$. Furthermore, the skew tunneling occurs in the presence of a finite total backreflection geometric phase, which requires the two interfaces of the barrier to be asymmetric ($\mu_L\neq\mu_R$). This acquired total backreflection geometric phase is valley-contrasted in the absence of the spin-orbit coupling, leading to a pure TVHE, and becomes spin-dependent in the presence of the spin-orbit coupling, which is the key factor in generating the TSHE.

The spin-orbit interactions are externally induced by the multilayer $\ce{WS2}$ substrate via the proximity effect, where we consider only the dominant valley-Zeeman term ($\lambda\nu s_z\sigma_0$) in our model. In fact, a small Rashba-type spin-orbit coupling $h_{R}=\lambda_R(\nu s_y\sigma_x-s_x\sigma_y)$ is also present. The DFT study suggests that both the valley-Zeeman coupling and the Rashba coupling are dependent on the interlayer distance of the multilayer $\ce{WS2}$ substrate with the value of $\lambda$ being much greater than that of $\lambda_R$ \cite{Yang_2016}. Consequently, the Rashba spin-orbit coupling is usually neglected when considering the transport in graphene tunnel junctions with proximity-induced spin-orbit interactions, such as in the case of the Andreev reflection in the proximitized graphene/superconductor junctions \cite{PhysRevB.108.134511,PhysRevB.108.024504}.

\begin{figure}[tp]
\begin{center}
\includegraphics[clip = true, width =\columnwidth]{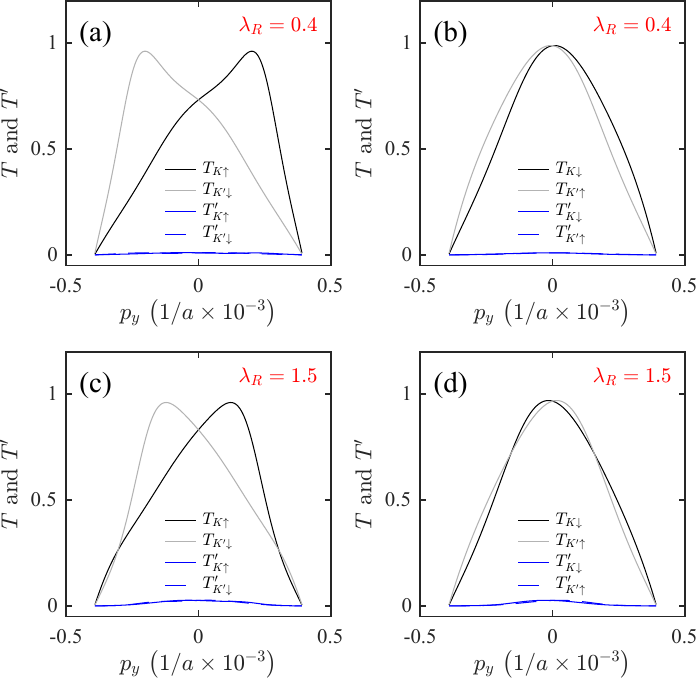}
\end{center}
\caption{The valley- and spin-dependent transmission probability versus the transverse momentum $p_y$ in the presence of the Rashba spin-orbit coupling. [(a), (b)] $\lambda_R=\SI{0.4}{\meV}$. [(c), (d)] $\lambda_R=\SI{1.5}{\meV}$. The other parameters are the same as that in Fig.\ \ref{fig:trans}.}
\label{fig:x}
\end{figure}

When considering the presence of the small Rashba spin-orbit coupling, the spin-flip scattering occurs because the Rashba term in the barrier region couples the spin-up and spin-down channels. The transmission amplitudes can be numerically obtained by the transfer matrix method. The scattering wave functions satisfy the following equation:
\begin{align}
    \mathcal{M}(\psi_{L,s_z}^++r\psi_{L,s_z}^-+r'\psi_{L,\bar{s}_z}^-)=t\psi_{R,s_z}^++t'\psi_{R,\bar{s}_z}^+,\label{eq:tmm}
\end{align}
where $s_z=\{\uparrow,\downarrow\}$ is the spin index. The basis states are given by
\begin{align}
    \psi_{j,\uparrow}^{\alpha}=\begin{pmatrix}
        \tilde\chi^{\alpha}_{j}\\\slashzero
    \end{pmatrix},\quad \psi_{j,\downarrow}^{\alpha}=\begin{pmatrix}
        \slashzero\\\tilde\chi^{\alpha}_{j}
    \end{pmatrix},
\end{align}
where $\alpha=+$ $(-)$ denotes the positive (negative) propagating direction, $j=L,R$ denote the left and right electrodes, respectively, $\slashzero$ is the $2\times1$ zero matrix, and $\tilde\chi^\alpha_j$ is the basis state given by Eqs.\ (\ref{eq:interface1}) and (\ref{eq:interface2}). $\mathcal{M}$ is the transfer matrix in the barrier region, which can be directly obtained by the integration of the secular equation $\bar{\mathcal{H}}_b\psi(x)=\epsilon\psi(x)$ \cite{PhysRevB.77.085423,PhysRevB.79.205428} with $\bar{\mathcal{H}}_b=\mathcal{H}_b+\lambda_R(\nu s_y\sigma_x-s_x\sigma_y)$ being the barrier Hamiltonian in the presence of the Rashba spin-orbit coupling, resulting in 
\begin{align}
    \mathcal{M}=e^{i\nu d s_0\sigma_x\mathcal{O}},\quad \mathcal{O}=\epsilon\times\mathds{1}-\bar{\mathcal{H}}_{b}|_{p_x=0}.
\end{align}
The amplitudes of the spin-conserving and spin-flip reflection (transmission) are denoted by $r$ and $r'$ ($t$ and $t'$), respectively, which can be obtained by solving Eq.\ (\ref{eq:tmm}). The spin-conserving transmission probability ($T=|t|^2$) and spin-flip transmission probability ($T'=|t'|^2$) are shown in Fig.\ \ref{fig:x}. It is shown that the probabilities of the spin-flip transmission induced by the Rashba spin-orbit coupling (blue lines) are significantly smaller than those of the spin-conserving transmission (black and gray lines). The tunneling is dominated by the spin-conserving process. For the finite valley index ($\nu$) and spin index ($s_z$), the transmission is still asymmetric, \textit{i.e.}, $T_{\nu s_z}(p_y)\neq T_{\nu s_z}(-p_y)$, which is responsible for the transverse valley and spin currents. Since the Rashba spin-orbit coupling preserves the time-reversal symmetry, the symmetric relation $T_{\nu s_z}(p_y)=T_{\bar\nu \bar{s}_z}(-p_y)$ is maintained, and thus the transverse charge current does not arise. Consequently, it is valid to neglect the small Rashba spin-orbit coupling in our model without influencing the main physics.

\section{Conclusion}\label{conc}

In summary, we propose a method for the generation of the TSHE and TVHE, which is based on the phase coherent tunneling in graphene heterojunctions. In the barrier region, the pseudospin staggered potential and the spin-orbit coupling are considered, which can be induced in graphene by the proximity effect of substrates. When the barrier of the tunnel junction has two asymmetric interfaces, the tunneling electrons acquire a finite spin- and valley-dependent geometric phase, leading to the spin- and valley-dependent skew coherent tunneling, which is responsible for the transverse spin and valley Hall currents. Distinct from the previously proposed TSHE and TVHE, this coherent-tunneling assisted Hall effect can be generated without the need for breaking time-reversal symmetry or in the absence of tilted Dirac cones, and exhibits large spin and valley Hall angles, suggesting potential applications for spintronic and valleytronic devices.

We note that, in the presence of an in-plane electric field, the spin, valley, and coupled spin-valley Hall effects can also be generated in graphene deposited on ferromagnetic substrates \cite{Dyrdał_2017} and monolayers of $\ce{MoS2}$ \cite{PhysRevLett.108.196802}, where the electron will acquire an anomalous transverse velocity proportional to the Berry curvature, giving rise to an intrinsic contribution to the Hall conductivity. The Hall effect is driven by the in-plane electric field and the Hall conductance can be obtained by the Kubo formula \cite{RevModPhys.82.1959,RevModPhys.82.1539}. However, in our model, we consider the electron transport in the ballistic regime, where the current is driven by the bias voltage across the tunnel junction and the conductance is obtained within the Landauer formalism \cite{datta1997electronic}. The predicted mechanism of the tunneling Hall effect is attribute to the coherence of the geometric phase acquired at the interfaces of the barrier and is different from that of the Hall effects proposed in Refs.\ \cite{Dyrdał_2017,PhysRevLett.108.196802}, where the Hall conductivity is associated with the acquired anomalous transverse velocities proportional to the Berry curvature in the presence of an in-plane electric field.


\begin{thebibliography}{47}%
\makeatletter
\providecommand \@ifxundefined [1]{%
 \@ifx{#1\undefined}
}%
\providecommand \@ifnum [1]{%
 \ifnum #1\expandafter \@firstoftwo
 \else \expandafter \@secondoftwo
 \fi
}%
\providecommand \@ifx [1]{%
 \ifx #1\expandafter \@firstoftwo
 \else \expandafter \@secondoftwo
 \fi
}%
\providecommand \natexlab [1]{#1}%
\providecommand \enquote  [1]{``#1''}%
\providecommand \bibnamefont  [1]{#1}%
\providecommand \bibfnamefont [1]{#1}%
\providecommand \citenamefont [1]{#1}%
\providecommand \href@noop [0]{\@secondoftwo}%
\providecommand \href [0]{\begingroup \@sanitize@url \@href}%
\providecommand \@href[1]{\@@startlink{#1}\@@href}%
\providecommand \@@href[1]{\endgroup#1\@@endlink}%
\providecommand \@sanitize@url [0]{\catcode `\\12\catcode `\$12\catcode
  `\&12\catcode `\#12\catcode `\^12\catcode `\_12\catcode `\%12\relax}%
\providecommand \@@startlink[1]{}%
\providecommand \@@endlink[0]{}%
\providecommand \url  [0]{\begingroup\@sanitize@url \@url }%
\providecommand \@url [1]{\endgroup\@href {#1}{\urlprefix }}%
\providecommand \urlprefix  [0]{URL }%
\providecommand \Eprint [0]{\href }%
\providecommand \doibase [0]{https://doi.org/}%
\providecommand \selectlanguage [0]{\@gobble}%
\providecommand \bibinfo  [0]{\@secondoftwo}%
\providecommand \bibfield  [0]{\@secondoftwo}%
\providecommand \translation [1]{[#1]}%
\providecommand \BibitemOpen [0]{}%
\providecommand \bibitemStop [0]{}%
\providecommand \bibitemNoStop [0]{.\EOS\space}%
\providecommand \EOS [0]{\spacefactor3000\relax}%
\providecommand \BibitemShut  [1]{\csname bibitem#1\endcsname}%
\let\auto@bib@innerbib\@empty
%</preamble>
\bibitem [{\citenamefont {\ifmmode \check{Z}\else
  \v{Z}\fi{}uti\ifmmode~\acute{c}\else \'{c}\fi{}}\ \emph
  {et~al.}(2004)\citenamefont {\ifmmode \check{Z}\else
  \v{Z}\fi{}uti\ifmmode~\acute{c}\else \'{c}\fi{}}, \citenamefont {Fabian},\
  and\ \citenamefont {Das~Sarma}}]{RevModPhys.76.323}%
  \BibitemOpen
  \bibfield  {author} {\bibinfo {author} {\bibfnamefont {I.}~\bibnamefont
  {\ifmmode \check{Z}\else \v{Z}\fi{}uti\ifmmode~\acute{c}\else \'{c}\fi{}}},
  \bibinfo {author} {\bibfnamefont {J.}~\bibnamefont {Fabian}},\ and\ \bibinfo
  {author} {\bibfnamefont {S.}~\bibnamefont {Das~Sarma}},\ }\bibfield  {title}
  {\bibinfo {title} {Spintronics: Fundamentals and applications},\ }\href
  {https://doi.org/10.1103/RevModPhys.76.323} {\bibfield  {journal} {\bibinfo
  {journal} {Rev. Mod. Phys.}\ }\textbf {\bibinfo {volume} {76}},\ \bibinfo
  {pages} {323} (\bibinfo {year} {2004})}\BibitemShut {NoStop}%
\bibitem [{\citenamefont {Wang}\ \emph {et~al.}(2018)\citenamefont {Wang},
  \citenamefont {Chernikov}, \citenamefont {Glazov}, \citenamefont {Heinz},
  \citenamefont {Marie}, \citenamefont {Amand},\ and\ \citenamefont
  {Urbaszek}}]{RevModPhys.90.021001}%
  \BibitemOpen
  \bibfield  {author} {\bibinfo {author} {\bibfnamefont {G.}~\bibnamefont
  {Wang}}, \bibinfo {author} {\bibfnamefont {A.}~\bibnamefont {Chernikov}},
  \bibinfo {author} {\bibfnamefont {M.~M.}\ \bibnamefont {Glazov}}, \bibinfo
  {author} {\bibfnamefont {T.~F.}\ \bibnamefont {Heinz}}, \bibinfo {author}
  {\bibfnamefont {X.}~\bibnamefont {Marie}}, \bibinfo {author} {\bibfnamefont
  {T.}~\bibnamefont {Amand}},\ and\ \bibinfo {author} {\bibfnamefont
  {B.}~\bibnamefont {Urbaszek}},\ }\bibfield  {title} {\bibinfo {title}
  {Colloquium: Excitons in atomically thin transition metal dichalcogenides},\
  }\href {https://doi.org/10.1103/RevModPhys.90.021001} {\bibfield  {journal}
  {\bibinfo  {journal} {Rev. Mod. Phys.}\ }\textbf {\bibinfo {volume} {90}},\
  \bibinfo {pages} {021001} (\bibinfo {year} {2018})}\BibitemShut {NoStop}%
\bibitem [{\citenamefont {Pulizzi}(2012)}]{pulizzi2012spintronics}%
  \BibitemOpen
  \bibfield  {author} {\bibinfo {author} {\bibfnamefont {F.}~\bibnamefont
  {Pulizzi}},\ }\bibfield  {title} {\bibinfo {title} {Spintronics},\ }\href
  {https://doi.org/10.1038/nmat3327} {\bibfield  {journal} {\bibinfo  {journal}
  {Nature materials}\ }\textbf {\bibinfo {volume} {11}},\ \bibinfo {pages}
  {367} (\bibinfo {year} {2012})}\BibitemShut {NoStop}%
\bibitem [{\citenamefont {Ferreira}\ \emph {et~al.}(2014)\citenamefont
  {Ferreira}, \citenamefont {Rappoport}, \citenamefont {Cazalilla},\ and\
  \citenamefont {Castro~Neto}}]{PhysRevLett.112.066601}%
  \BibitemOpen
  \bibfield  {author} {\bibinfo {author} {\bibfnamefont {A.}~\bibnamefont
  {Ferreira}}, \bibinfo {author} {\bibfnamefont {T.~G.}\ \bibnamefont
  {Rappoport}}, \bibinfo {author} {\bibfnamefont {M.~A.}\ \bibnamefont
  {Cazalilla}},\ and\ \bibinfo {author} {\bibfnamefont {A.~H.}\ \bibnamefont
  {Castro~Neto}},\ }\bibfield  {title} {\bibinfo {title} {Extrinsic spin {Hall}
  effect induced by resonant skew scattering in graphene},\ }\href
  {https://doi.org/10.1103/PhysRevLett.112.066601} {\bibfield  {journal}
  {\bibinfo  {journal} {Phys. Rev. Lett.}\ }\textbf {\bibinfo {volume} {112}},\
  \bibinfo {pages} {066601} (\bibinfo {year} {2014})}\BibitemShut {NoStop}%
\bibitem [{\citenamefont {Zhang}(2000)}]{PhysRevLett.85.393}%
  \BibitemOpen
  \bibfield  {author} {\bibinfo {author} {\bibfnamefont {S.}~\bibnamefont
  {Zhang}},\ }\bibfield  {title} {\bibinfo {title} {Spin {Hall} effect in the
  presence of spin diffusion},\ }\href
  {https://doi.org/10.1103/PhysRevLett.85.393} {\bibfield  {journal} {\bibinfo
  {journal} {Phys. Rev. Lett.}\ }\textbf {\bibinfo {volume} {85}},\ \bibinfo
  {pages} {393} (\bibinfo {year} {2000})}\BibitemShut {NoStop}%
\bibitem [{\citenamefont {Sinova}\ \emph {et~al.}(2004)\citenamefont {Sinova},
  \citenamefont {Culcer}, \citenamefont {Niu}, \citenamefont {Sinitsyn},
  \citenamefont {Jungwirth},\ and\ \citenamefont
  {MacDonald}}]{PhysRevLett.92.126603}%
  \BibitemOpen
  \bibfield  {author} {\bibinfo {author} {\bibfnamefont {J.}~\bibnamefont
  {Sinova}}, \bibinfo {author} {\bibfnamefont {D.}~\bibnamefont {Culcer}},
  \bibinfo {author} {\bibfnamefont {Q.}~\bibnamefont {Niu}}, \bibinfo {author}
  {\bibfnamefont {N.~A.}\ \bibnamefont {Sinitsyn}}, \bibinfo {author}
  {\bibfnamefont {T.}~\bibnamefont {Jungwirth}},\ and\ \bibinfo {author}
  {\bibfnamefont {A.~H.}\ \bibnamefont {MacDonald}},\ }\bibfield  {title}
  {\bibinfo {title} {Universal intrinsic spin {Hall} effect},\ }\href
  {https://doi.org/10.1103/PhysRevLett.92.126603} {\bibfield  {journal}
  {\bibinfo  {journal} {Phys. Rev. Lett.}\ }\textbf {\bibinfo {volume} {92}},\
  \bibinfo {pages} {126603} (\bibinfo {year} {2004})}\BibitemShut {NoStop}%
\bibitem [{\citenamefont {Guo}\ \emph {et~al.}(2008)\citenamefont {Guo},
  \citenamefont {Murakami}, \citenamefont {Chen},\ and\ \citenamefont
  {Nagaosa}}]{PhysRevLett.100.096401}%
  \BibitemOpen
  \bibfield  {author} {\bibinfo {author} {\bibfnamefont {G.~Y.}\ \bibnamefont
  {Guo}}, \bibinfo {author} {\bibfnamefont {S.}~\bibnamefont {Murakami}},
  \bibinfo {author} {\bibfnamefont {T.-W.}\ \bibnamefont {Chen}},\ and\
  \bibinfo {author} {\bibfnamefont {N.}~\bibnamefont {Nagaosa}},\ }\bibfield
  {title} {\bibinfo {title} {Intrinsic spin {Hall} effect in platinum:
  First-principles calculations},\ }\href
  {https://doi.org/10.1103/PhysRevLett.100.096401} {\bibfield  {journal}
  {\bibinfo  {journal} {Phys. Rev. Lett.}\ }\textbf {\bibinfo {volume} {100}},\
  \bibinfo {pages} {096401} (\bibinfo {year} {2008})}\BibitemShut {NoStop}%
\bibitem [{\citenamefont {Qi}\ \emph {et~al.}(2015)\citenamefont {Qi},
  \citenamefont {Li}, \citenamefont {Niu},\ and\ \citenamefont
  {Feng}}]{PhysRevB.92.121403}%
  \BibitemOpen
  \bibfield  {author} {\bibinfo {author} {\bibfnamefont {J.}~\bibnamefont
  {Qi}}, \bibinfo {author} {\bibfnamefont {X.}~\bibnamefont {Li}}, \bibinfo
  {author} {\bibfnamefont {Q.}~\bibnamefont {Niu}},\ and\ \bibinfo {author}
  {\bibfnamefont {J.}~\bibnamefont {Feng}},\ }\bibfield  {title} {\bibinfo
  {title} {Giant and tunable valley degeneracy splitting in
  {${\mathrm{MoTe}}_{2}$}},\ }\href
  {https://doi.org/10.1103/PhysRevB.92.121403} {\bibfield  {journal} {\bibinfo
  {journal} {Phys. Rev. B}\ }\textbf {\bibinfo {volume} {92}},\ \bibinfo
  {pages} {121403} (\bibinfo {year} {2015})}\BibitemShut {NoStop}%
\bibitem [{\citenamefont {Mak}\ \emph {et~al.}(2014)\citenamefont {Mak},
  \citenamefont {McGill}, \citenamefont {Park},\ and\ \citenamefont
  {McEuen}}]{mak2014valley}%
  \BibitemOpen
  \bibfield  {author} {\bibinfo {author} {\bibfnamefont {K.~F.}\ \bibnamefont
  {Mak}}, \bibinfo {author} {\bibfnamefont {K.~L.}\ \bibnamefont {McGill}},
  \bibinfo {author} {\bibfnamefont {J.}~\bibnamefont {Park}},\ and\ \bibinfo
  {author} {\bibfnamefont {P.~L.}\ \bibnamefont {McEuen}},\ }\bibfield  {title}
  {\bibinfo {title} {The valley {Hall} effect in {${\mathrm{MoS}}_{2}$}
  transistors},\ }\href
  {https://www.science.org/doi/abs/10.1126/science.1250140} {\bibfield
  {journal} {\bibinfo  {journal} {Science}\ }\textbf {\bibinfo {volume}
  {344}},\ \bibinfo {pages} {1489} (\bibinfo {year} {2014})}\BibitemShut
  {NoStop}%
\bibitem [{\citenamefont {Lundt}\ \emph {et~al.}(2019)\citenamefont {Lundt},
  \citenamefont {Dusanowski}, \citenamefont {Sedov}, \citenamefont {Stepanov},
  \citenamefont {Glazov}, \citenamefont {Klembt}, \citenamefont {Klaas},
  \citenamefont {Beierlein}, \citenamefont {Qin}, \citenamefont {Tongay} \emph
  {et~al.}}]{lundt2019optical}%
  \BibitemOpen
  \bibfield  {author} {\bibinfo {author} {\bibfnamefont {N.}~\bibnamefont
  {Lundt}}, \bibinfo {author} {\bibfnamefont {{\L}.}~\bibnamefont
  {Dusanowski}}, \bibinfo {author} {\bibfnamefont {E.}~\bibnamefont {Sedov}},
  \bibinfo {author} {\bibfnamefont {P.}~\bibnamefont {Stepanov}}, \bibinfo
  {author} {\bibfnamefont {M.~M.}\ \bibnamefont {Glazov}}, \bibinfo {author}
  {\bibfnamefont {S.}~\bibnamefont {Klembt}}, \bibinfo {author} {\bibfnamefont
  {M.}~\bibnamefont {Klaas}}, \bibinfo {author} {\bibfnamefont
  {J.}~\bibnamefont {Beierlein}}, \bibinfo {author} {\bibfnamefont
  {Y.}~\bibnamefont {Qin}}, \bibinfo {author} {\bibfnamefont {S.}~\bibnamefont
  {Tongay}}, \emph {et~al.},\ }\bibfield  {title} {\bibinfo {title} {Optical
  valley {Hall} effect for highly valley-coherent exciton-polaritons in an
  atomically thin semiconductor},\ }\href
  {https://doi.org/10.1038/s41565-019-0492-0} {\bibfield  {journal} {\bibinfo
  {journal} {Nature nanotechnology}\ }\textbf {\bibinfo {volume} {14}},\
  \bibinfo {pages} {770} (\bibinfo {year} {2019})}\BibitemShut {NoStop}%
\bibitem [{\citenamefont {Xiao}\ \emph {et~al.}(2007)\citenamefont {Xiao},
  \citenamefont {Yao},\ and\ \citenamefont {Niu}}]{PhysRevLett.99.236809}%
  \BibitemOpen
  \bibfield  {author} {\bibinfo {author} {\bibfnamefont {D.}~\bibnamefont
  {Xiao}}, \bibinfo {author} {\bibfnamefont {W.}~\bibnamefont {Yao}},\ and\
  \bibinfo {author} {\bibfnamefont {Q.}~\bibnamefont {Niu}},\ }\bibfield
  {title} {\bibinfo {title} {Valley-contrasting physics in graphene: Magnetic
  moment and topological transport},\ }\href
  {https://doi.org/10.1103/PhysRevLett.99.236809} {\bibfield  {journal}
  {\bibinfo  {journal} {Phys. Rev. Lett.}\ }\textbf {\bibinfo {volume} {99}},\
  \bibinfo {pages} {236809} (\bibinfo {year} {2007})}\BibitemShut {NoStop}%
\bibitem [{\citenamefont {Chang}\ and\ \citenamefont
  {Niu}(1995)}]{PhysRevLett.75.1348}%
  \BibitemOpen
  \bibfield  {author} {\bibinfo {author} {\bibfnamefont {M.-C.}\ \bibnamefont
  {Chang}}\ and\ \bibinfo {author} {\bibfnamefont {Q.}~\bibnamefont {Niu}},\
  }\bibfield  {title} {\bibinfo {title} {Berry phase, hyperorbits, and the
  {Hofstadter} spectrum},\ }\href {https://doi.org/10.1103/PhysRevLett.75.1348}
  {\bibfield  {journal} {\bibinfo  {journal} {Phys. Rev. Lett.}\ }\textbf
  {\bibinfo {volume} {75}},\ \bibinfo {pages} {1348} (\bibinfo {year}
  {1995})}\BibitemShut {NoStop}%
\bibitem [{\citenamefont {Xiao}\ \emph {et~al.}(2012)\citenamefont {Xiao},
  \citenamefont {Liu}, \citenamefont {Feng}, \citenamefont {Xu},\ and\
  \citenamefont {Yao}}]{PhysRevLett.108.196802}%
  \BibitemOpen
  \bibfield  {author} {\bibinfo {author} {\bibfnamefont {D.}~\bibnamefont
  {Xiao}}, \bibinfo {author} {\bibfnamefont {G.-B.}\ \bibnamefont {Liu}},
  \bibinfo {author} {\bibfnamefont {W.}~\bibnamefont {Feng}}, \bibinfo {author}
  {\bibfnamefont {X.}~\bibnamefont {Xu}},\ and\ \bibinfo {author}
  {\bibfnamefont {W.}~\bibnamefont {Yao}},\ }\bibfield  {title} {\bibinfo
  {title} {Coupled spin and valley physics in monolayers of
  {${\mathrm{MoS}}_{2}$} and other group-{VI} dichalcogenides},\ }\href
  {https://doi.org/10.1103/PhysRevLett.108.196802} {\bibfield  {journal}
  {\bibinfo  {journal} {Phys. Rev. Lett.}\ }\textbf {\bibinfo {volume} {108}},\
  \bibinfo {pages} {196802} (\bibinfo {year} {2012})}\BibitemShut {NoStop}%
\bibitem [{\citenamefont {Ominato}\ \emph {et~al.}(2020)\citenamefont
  {Ominato}, \citenamefont {Fujimoto},\ and\ \citenamefont
  {Matsuo}}]{PhysRevLett.124.166803}%
  \BibitemOpen
  \bibfield  {author} {\bibinfo {author} {\bibfnamefont {Y.}~\bibnamefont
  {Ominato}}, \bibinfo {author} {\bibfnamefont {J.}~\bibnamefont {Fujimoto}},\
  and\ \bibinfo {author} {\bibfnamefont {M.}~\bibnamefont {Matsuo}},\
  }\bibfield  {title} {\bibinfo {title} {Valley-dependent spin transport in
  monolayer transition-metal dichalcogenides},\ }\href
  {https://doi.org/10.1103/PhysRevLett.124.166803} {\bibfield  {journal}
  {\bibinfo  {journal} {Phys. Rev. Lett.}\ }\textbf {\bibinfo {volume} {124}},\
  \bibinfo {pages} {166803} (\bibinfo {year} {2020})}\BibitemShut {NoStop}%
\bibitem [{\citenamefont {Matos-Abiague}\ and\ \citenamefont
  {Fabian}(2015)}]{PhysRevLett.115.056602}%
  \BibitemOpen
  \bibfield  {author} {\bibinfo {author} {\bibfnamefont {A.}~\bibnamefont
  {Matos-Abiague}}\ and\ \bibinfo {author} {\bibfnamefont {J.}~\bibnamefont
  {Fabian}},\ }\bibfield  {title} {\bibinfo {title} {Tunneling anomalous and
  spin {Hall} effects},\ }\href
  {https://doi.org/10.1103/PhysRevLett.115.056602} {\bibfield  {journal}
  {\bibinfo  {journal} {Phys. Rev. Lett.}\ }\textbf {\bibinfo {volume} {115}},\
  \bibinfo {pages} {056602} (\bibinfo {year} {2015})}\BibitemShut {NoStop}%
\bibitem [{\citenamefont {Vedyayev}\ \emph {et~al.}(2013)\citenamefont
  {Vedyayev}, \citenamefont {Ryzhanova}, \citenamefont {Strelkov},\ and\
  \citenamefont {Dieny}}]{PhysRevLett.110.247204}%
  \BibitemOpen
  \bibfield  {author} {\bibinfo {author} {\bibfnamefont {A.}~\bibnamefont
  {Vedyayev}}, \bibinfo {author} {\bibfnamefont {N.}~\bibnamefont {Ryzhanova}},
  \bibinfo {author} {\bibfnamefont {N.}~\bibnamefont {Strelkov}},\ and\
  \bibinfo {author} {\bibfnamefont {B.}~\bibnamefont {Dieny}},\ }\bibfield
  {title} {\bibinfo {title} {Spontaneous anomalous and spin {Hall} effects due
  to spin-orbit scattering of evanescent wave functions in magnetic tunnel
  junctions},\ }\href {https://doi.org/10.1103/PhysRevLett.110.247204}
  {\bibfield  {journal} {\bibinfo  {journal} {Phys. Rev. Lett.}\ }\textbf
  {\bibinfo {volume} {110}},\ \bibinfo {pages} {247204} (\bibinfo {year}
  {2013})}\BibitemShut {NoStop}%
\bibitem [{\citenamefont {Scharf}\ \emph {et~al.}(2016)\citenamefont {Scharf},
  \citenamefont {Matos-Abiague}, \citenamefont {Han}, \citenamefont
  {Hankiewicz},\ and\ \citenamefont {\ifmmode \check{Z}\else
  \v{Z}\fi{}uti\ifmmode~\acute{c}\else \'{c}\fi{}}}]{PhysRevLett.117.166806}%
  \BibitemOpen
  \bibfield  {author} {\bibinfo {author} {\bibfnamefont {B.}~\bibnamefont
  {Scharf}}, \bibinfo {author} {\bibfnamefont {A.}~\bibnamefont
  {Matos-Abiague}}, \bibinfo {author} {\bibfnamefont {J.~E.}\ \bibnamefont
  {Han}}, \bibinfo {author} {\bibfnamefont {E.~M.}\ \bibnamefont
  {Hankiewicz}},\ and\ \bibinfo {author} {\bibfnamefont {I.}~\bibnamefont
  {\ifmmode \check{Z}\else \v{Z}\fi{}uti\ifmmode~\acute{c}\else \'{c}\fi{}}},\
  }\bibfield  {title} {\bibinfo {title} {Tunneling planar {Hall} effect in
  topological insulators: Spin valves and amplifiers},\ }\href
  {https://doi.org/10.1103/PhysRevLett.117.166806} {\bibfield  {journal}
  {\bibinfo  {journal} {Phys. Rev. Lett.}\ }\textbf {\bibinfo {volume} {117}},\
  \bibinfo {pages} {166806} (\bibinfo {year} {2016})}\BibitemShut {NoStop}%
\bibitem [{\citenamefont {Zhang}\ \emph {et~al.}(2023)\citenamefont {Zhang},
  \citenamefont {Shao}, \citenamefont {Wang}, \citenamefont {Yang},
  \citenamefont {Yang},\ and\ \citenamefont
  {Tsymbal}}]{PhysRevLett.131.246301}%
  \BibitemOpen
  \bibfield  {author} {\bibinfo {author} {\bibfnamefont {S.-H.}\ \bibnamefont
  {Zhang}}, \bibinfo {author} {\bibfnamefont {D.-F.}\ \bibnamefont {Shao}},
  \bibinfo {author} {\bibfnamefont {Z.-A.}\ \bibnamefont {Wang}}, \bibinfo
  {author} {\bibfnamefont {J.}~\bibnamefont {Yang}}, \bibinfo {author}
  {\bibfnamefont {W.}~\bibnamefont {Yang}},\ and\ \bibinfo {author}
  {\bibfnamefont {E.~Y.}\ \bibnamefont {Tsymbal}},\ }\bibfield  {title}
  {\bibinfo {title} {Tunneling valley {Hall} effect driven by tilted {Dirac}
  fermions},\ }\href {https://doi.org/10.1103/PhysRevLett.131.246301}
  {\bibfield  {journal} {\bibinfo  {journal} {Phys. Rev. Lett.}\ }\textbf
  {\bibinfo {volume} {131}},\ \bibinfo {pages} {246301} (\bibinfo {year}
  {2023})}\BibitemShut {NoStop}%
\bibitem [{\citenamefont {Zeng}(2024)}]{PhysRevB.110.024511}%
  \BibitemOpen
  \bibfield  {author} {\bibinfo {author} {\bibfnamefont {W.}~\bibnamefont
  {Zeng}},\ }\bibfield  {title} {\bibinfo {title} {Tunneling chirality {Hall}
  effect in type-{I} {Weyl} semimetals},\ }\href
  {https://doi.org/10.1103/PhysRevB.110.024511} {\bibfield  {journal} {\bibinfo
   {journal} {Phys. Rev. B}\ }\textbf {\bibinfo {volume} {110}},\ \bibinfo
  {pages} {024511} (\bibinfo {year} {2024})}\BibitemShut {NoStop}%
\bibitem [{\citenamefont {Beenakker}(2008)}]{RevModPhys.80.1337}%
  \BibitemOpen
  \bibfield  {author} {\bibinfo {author} {\bibfnamefont {C.~W.~J.}\
  \bibnamefont {Beenakker}},\ }\bibfield  {title} {\bibinfo {title}
  {Colloquium: Andreev reflection and klein tunneling in graphene},\ }\href
  {https://doi.org/10.1103/RevModPhys.80.1337} {\bibfield  {journal} {\bibinfo
  {journal} {Rev. Mod. Phys.}\ }\textbf {\bibinfo {volume} {80}},\ \bibinfo
  {pages} {1337} (\bibinfo {year} {2008})}\BibitemShut {NoStop}%
\bibitem [{\citenamefont {Zihlmann}\ \emph {et~al.}(2018)\citenamefont
  {Zihlmann}, \citenamefont {Cummings}, \citenamefont {Garcia}, \citenamefont
  {Kedves}, \citenamefont {Watanabe}, \citenamefont {Taniguchi}, \citenamefont
  {Sch\"onenberger},\ and\ \citenamefont {Makk}}]{PhysRevB.97.075434}%
  \BibitemOpen
  \bibfield  {author} {\bibinfo {author} {\bibfnamefont {S.}~\bibnamefont
  {Zihlmann}}, \bibinfo {author} {\bibfnamefont {A.~W.}\ \bibnamefont
  {Cummings}}, \bibinfo {author} {\bibfnamefont {J.~H.}\ \bibnamefont
  {Garcia}}, \bibinfo {author} {\bibfnamefont {M.}~\bibnamefont {Kedves}},
  \bibinfo {author} {\bibfnamefont {K.}~\bibnamefont {Watanabe}}, \bibinfo
  {author} {\bibfnamefont {T.}~\bibnamefont {Taniguchi}}, \bibinfo {author}
  {\bibfnamefont {C.}~\bibnamefont {Sch\"onenberger}},\ and\ \bibinfo {author}
  {\bibfnamefont {P.}~\bibnamefont {Makk}},\ }\bibfield  {title} {\bibinfo
  {title} {Large spin relaxation anisotropy and valley-{Zeeman} spin-orbit
  coupling in {${\mathrm{WSe}}_{2}$/graphene/$h$-BN} heterostructures},\ }\href
  {https://doi.org/10.1103/PhysRevB.97.075434} {\bibfield  {journal} {\bibinfo
  {journal} {Phys. Rev. B}\ }\textbf {\bibinfo {volume} {97}},\ \bibinfo
  {pages} {075434} (\bibinfo {year} {2018})}\BibitemShut {NoStop}%
\bibitem [{\citenamefont {Zollner}\ and\ \citenamefont
  {Fabian}(2022)}]{PhysRevB.106.035137}%
  \BibitemOpen
  \bibfield  {author} {\bibinfo {author} {\bibfnamefont {K.}~\bibnamefont
  {Zollner}}\ and\ \bibinfo {author} {\bibfnamefont {J.}~\bibnamefont
  {Fabian}},\ }\bibfield  {title} {\bibinfo {title} {Proximity effects in
  graphene on monolayers of transition-metal phosphorus trichalcogenides
  {$M\mathrm{P}{X}_{3}$ $(M:\mathrm{Mn}, \mathrm{Fe}, \mathrm{Ni}, \mathrm{Co},
  \mathrm{and} X: \mathrm{S}, \mathrm{Se})$}},\ }\href
  {https://doi.org/10.1103/PhysRevB.106.035137} {\bibfield  {journal} {\bibinfo
   {journal} {Phys. Rev. B}\ }\textbf {\bibinfo {volume} {106}},\ \bibinfo
  {pages} {035137} (\bibinfo {year} {2022})}\BibitemShut {NoStop}%
\bibitem [{\citenamefont {Khatibi}\ and\ \citenamefont
  {Power}(2022)}]{PhysRevB.106.125417}%
  \BibitemOpen
  \bibfield  {author} {\bibinfo {author} {\bibfnamefont {Z.}~\bibnamefont
  {Khatibi}}\ and\ \bibinfo {author} {\bibfnamefont {S.~R.}\ \bibnamefont
  {Power}},\ }\bibfield  {title} {\bibinfo {title} {Proximity spin-orbit
  coupling in graphene on alloyed transition metal dichalcogenides},\ }\href
  {https://doi.org/10.1103/PhysRevB.106.125417} {\bibfield  {journal} {\bibinfo
   {journal} {Phys. Rev. B}\ }\textbf {\bibinfo {volume} {106}},\ \bibinfo
  {pages} {125417} (\bibinfo {year} {2022})}\BibitemShut {NoStop}%
\bibitem [{\citenamefont {Wang}\ \emph {et~al.}(2015)\citenamefont {Wang},
  \citenamefont {Ki}, \citenamefont {Chen}, \citenamefont {Berger},
  \citenamefont {MacDonald},\ and\ \citenamefont {Morpurgo}}]{wang2015strong}%
  \BibitemOpen
  \bibfield  {author} {\bibinfo {author} {\bibfnamefont {Z.}~\bibnamefont
  {Wang}}, \bibinfo {author} {\bibfnamefont {D.-K.}\ \bibnamefont {Ki}},
  \bibinfo {author} {\bibfnamefont {H.}~\bibnamefont {Chen}}, \bibinfo {author}
  {\bibfnamefont {H.}~\bibnamefont {Berger}}, \bibinfo {author} {\bibfnamefont
  {A.~H.}\ \bibnamefont {MacDonald}},\ and\ \bibinfo {author} {\bibfnamefont
  {A.~F.}\ \bibnamefont {Morpurgo}},\ }\bibfield  {title} {\bibinfo {title}
  {Strong interface-induced spin--orbit interaction in graphene on
  {${\mathrm{WS}}_{2}$}},\ }\href {https://doi.org/10.1038/ncomms9339}
  {\bibfield  {journal} {\bibinfo  {journal} {Nature communications}\ }\textbf
  {\bibinfo {volume} {6}},\ \bibinfo {pages} {8339} (\bibinfo {year}
  {2015})}\BibitemShut {NoStop}%
\bibitem [{\citenamefont {Zubair}\ \emph {et~al.}(2020)\citenamefont {Zubair},
  \citenamefont {Vasilopoulos},\ and\ \citenamefont
  {Tahir}}]{PhysRevB.101.165436}%
  \BibitemOpen
  \bibfield  {author} {\bibinfo {author} {\bibfnamefont {M.}~\bibnamefont
  {Zubair}}, \bibinfo {author} {\bibfnamefont {P.}~\bibnamefont
  {Vasilopoulos}},\ and\ \bibinfo {author} {\bibfnamefont {M.}~\bibnamefont
  {Tahir}},\ }\bibfield  {title} {\bibinfo {title} {Influence of interface
  induced {valley-Zeeman} and spin-orbit couplings on transport in
  heterostructures of graphene on {${\mathrm{WSe}}_{2}$}},\ }\href
  {https://doi.org/10.1103/PhysRevB.101.165436} {\bibfield  {journal} {\bibinfo
   {journal} {Phys. Rev. B}\ }\textbf {\bibinfo {volume} {101}},\ \bibinfo
  {pages} {165436} (\bibinfo {year} {2020})}\BibitemShut {NoStop}%
\bibitem [{\citenamefont {Yang}\ \emph {et~al.}(2016)\citenamefont {Yang},
  \citenamefont {Tu}, \citenamefont {Kim}, \citenamefont {Wu}, \citenamefont
  {Wang}, \citenamefont {Alicea}, \citenamefont {Wu}, \citenamefont
  {Bockrath},\ and\ \citenamefont {Shi}}]{Yang_2016}%
  \BibitemOpen
  \bibfield  {author} {\bibinfo {author} {\bibfnamefont {B.}~\bibnamefont
  {Yang}}, \bibinfo {author} {\bibfnamefont {M.-F.}\ \bibnamefont {Tu}},
  \bibinfo {author} {\bibfnamefont {J.}~\bibnamefont {Kim}}, \bibinfo {author}
  {\bibfnamefont {Y.}~\bibnamefont {Wu}}, \bibinfo {author} {\bibfnamefont
  {H.}~\bibnamefont {Wang}}, \bibinfo {author} {\bibfnamefont {J.}~\bibnamefont
  {Alicea}}, \bibinfo {author} {\bibfnamefont {R.}~\bibnamefont {Wu}}, \bibinfo
  {author} {\bibfnamefont {M.}~\bibnamefont {Bockrath}},\ and\ \bibinfo
  {author} {\bibfnamefont {J.}~\bibnamefont {Shi}},\ }\bibfield  {title}
  {\bibinfo {title} {Tunable spin–orbit coupling and symmetry-protected edge
  states in {graphene/$\mathrm{WS}_2$}},\ }\href
  {https://doi.org/10.1088/2053-1583/3/3/031012} {\bibfield  {journal}
  {\bibinfo  {journal} {2D Materials}\ }\textbf {\bibinfo {volume} {3}},\
  \bibinfo {pages} {031012} (\bibinfo {year} {2016})}\BibitemShut {NoStop}%
\bibitem [{\citenamefont {Cheng}\ and\ \citenamefont
  {Sun}(2022)}]{PhysRevB.105.165427}%
  \BibitemOpen
  \bibfield  {author} {\bibinfo {author} {\bibfnamefont {Q.}~\bibnamefont
  {Cheng}}\ and\ \bibinfo {author} {\bibfnamefont {Q.-F.}\ \bibnamefont
  {Sun}},\ }\bibfield  {title} {\bibinfo {title} {Spin-valley-resolved energy
  spectra of quantum dots in the graphene/transition metal dichalcogenides
  system},\ }\href {https://doi.org/10.1103/PhysRevB.105.165427} {\bibfield
  {journal} {\bibinfo  {journal} {Phys. Rev. B}\ }\textbf {\bibinfo {volume}
  {105}},\ \bibinfo {pages} {165427} (\bibinfo {year} {2022})}\BibitemShut
  {NoStop}%
\bibitem [{\citenamefont {Zhao}\ \emph {et~al.}(2023)\citenamefont {Zhao},
  \citenamefont {Gao}, \citenamefont {Cheng},\ and\ \citenamefont
  {Sun}}]{PhysRevB.108.134511}%
  \BibitemOpen
  \bibfield  {author} {\bibinfo {author} {\bibfnamefont {S.-C.}\ \bibnamefont
  {Zhao}}, \bibinfo {author} {\bibfnamefont {L.}~\bibnamefont {Gao}}, \bibinfo
  {author} {\bibfnamefont {Q.}~\bibnamefont {Cheng}},\ and\ \bibinfo {author}
  {\bibfnamefont {Q.-F.}\ \bibnamefont {Sun}},\ }\bibfield  {title} {\bibinfo
  {title} {Perfect crossed {Andreev} reflection in the proximitized
  graphene/superconductor/proximitized graphene junctions},\ }\href
  {https://doi.org/10.1103/PhysRevB.108.134511} {\bibfield  {journal} {\bibinfo
   {journal} {Phys. Rev. B}\ }\textbf {\bibinfo {volume} {108}},\ \bibinfo
  {pages} {134511} (\bibinfo {year} {2023})}\BibitemShut {NoStop}%
\bibitem [{\citenamefont {Datta}(1997)}]{datta1997electronic}%
  \BibitemOpen
  \bibfield  {author} {\bibinfo {author} {\bibfnamefont {S.}~\bibnamefont
  {Datta}},\ }\href@noop {} {\emph {\bibinfo {title} {Electronic transport in
  mesoscopic systems}}}\ (\bibinfo  {publisher} {Cambridge university press},\
  \bibinfo {year} {1997})\BibitemShut {NoStop}%
\bibitem [{\citenamefont {Shytov}\ \emph {et~al.}(2008)\citenamefont {Shytov},
  \citenamefont {Rudner},\ and\ \citenamefont
  {Levitov}}]{PhysRevLett.101.156804}%
  \BibitemOpen
  \bibfield  {author} {\bibinfo {author} {\bibfnamefont {A.~V.}\ \bibnamefont
  {Shytov}}, \bibinfo {author} {\bibfnamefont {M.~S.}\ \bibnamefont {Rudner}},\
  and\ \bibinfo {author} {\bibfnamefont {L.~S.}\ \bibnamefont {Levitov}},\
  }\bibfield  {title} {\bibinfo {title} {Klein backscattering and
  {Fabry-P\'erot} interference in graphene heterojunctions},\ }\href
  {https://doi.org/10.1103/PhysRevLett.101.156804} {\bibfield  {journal}
  {\bibinfo  {journal} {Phys. Rev. Lett.}\ }\textbf {\bibinfo {volume} {101}},\
  \bibinfo {pages} {156804} (\bibinfo {year} {2008})}\BibitemShut {NoStop}%
\bibitem [{\citenamefont {Mead}(1992)}]{RevModPhys.64.51}%
  \BibitemOpen
  \bibfield  {author} {\bibinfo {author} {\bibfnamefont {C.~A.}\ \bibnamefont
  {Mead}},\ }\bibfield  {title} {\bibinfo {title} {The geometric phase in
  molecular systems},\ }\href {https://doi.org/10.1103/RevModPhys.64.51}
  {\bibfield  {journal} {\bibinfo  {journal} {Rev. Mod. Phys.}\ }\textbf
  {\bibinfo {volume} {64}},\ \bibinfo {pages} {51} (\bibinfo {year}
  {1992})}\BibitemShut {NoStop}%
\bibitem [{\citenamefont {Resta}(1994)}]{RevModPhys.66.899}%
  \BibitemOpen
  \bibfield  {author} {\bibinfo {author} {\bibfnamefont {R.}~\bibnamefont
  {Resta}},\ }\bibfield  {title} {\bibinfo {title} {Macroscopic polarization in
  crystalline dielectrics: the geometric phase approach},\ }\href
  {https://doi.org/10.1103/RevModPhys.66.899} {\bibfield  {journal} {\bibinfo
  {journal} {Rev. Mod. Phys.}\ }\textbf {\bibinfo {volume} {66}},\ \bibinfo
  {pages} {899} (\bibinfo {year} {1994})}\BibitemShut {NoStop}%
\bibitem [{\citenamefont {Anandan}(1992)}]{anandan1992geometric}%
  \BibitemOpen
  \bibfield  {author} {\bibinfo {author} {\bibfnamefont {J.}~\bibnamefont
  {Anandan}},\ }\bibfield  {title} {\bibinfo {title} {The geometric phase},\
  }\href {https://doi.org/10.1038/360307a0} {\bibfield  {journal} {\bibinfo
  {journal} {Nature}\ }\textbf {\bibinfo {volume} {360}},\ \bibinfo {pages}
  {307} (\bibinfo {year} {1992})}\BibitemShut {NoStop}%
\bibitem [{\citenamefont {Choi}\ \emph {et~al.}(2013)\citenamefont {Choi},
  \citenamefont {Park},\ and\ \citenamefont {Sim}}]{PhysRevB.87.165420}%
  \BibitemOpen
  \bibfield  {author} {\bibinfo {author} {\bibfnamefont {S.-J.}\ \bibnamefont
  {Choi}}, \bibinfo {author} {\bibfnamefont {S.}~\bibnamefont {Park}},\ and\
  \bibinfo {author} {\bibfnamefont {H.-S.}\ \bibnamefont {Sim}},\ }\bibfield
  {title} {\bibinfo {title} {Tunable geometric phase of {Dirac} fermions in a
  topological junction},\ }\href {https://doi.org/10.1103/PhysRevB.87.165420}
  {\bibfield  {journal} {\bibinfo  {journal} {Phys. Rev. B}\ }\textbf {\bibinfo
  {volume} {87}},\ \bibinfo {pages} {165420} (\bibinfo {year}
  {2013})}\BibitemShut {NoStop}%
\bibitem [{\citenamefont {Cormann}\ \emph {et~al.}(2016)\citenamefont
  {Cormann}, \citenamefont {Remy}, \citenamefont {Kolaric},\ and\ \citenamefont
  {Caudano}}]{PhysRevA.93.042124}%
  \BibitemOpen
  \bibfield  {author} {\bibinfo {author} {\bibfnamefont {M.}~\bibnamefont
  {Cormann}}, \bibinfo {author} {\bibfnamefont {M.}~\bibnamefont {Remy}},
  \bibinfo {author} {\bibfnamefont {B.}~\bibnamefont {Kolaric}},\ and\ \bibinfo
  {author} {\bibfnamefont {Y.}~\bibnamefont {Caudano}},\ }\bibfield  {title}
  {\bibinfo {title} {Revealing geometric phases in modular and weak values with
  a quantum eraser},\ }\href {https://doi.org/10.1103/PhysRevA.93.042124}
  {\bibfield  {journal} {\bibinfo  {journal} {Phys. Rev. A}\ }\textbf {\bibinfo
  {volume} {93}},\ \bibinfo {pages} {042124} (\bibinfo {year}
  {2016})}\BibitemShut {NoStop}%
\bibitem [{\citenamefont {Choi}\ \emph {et~al.}(2014)\citenamefont {Choi},
  \citenamefont {Park},\ and\ \citenamefont {Sim}}]{PhysRevB.89.155412}%
  \BibitemOpen
  \bibfield  {author} {\bibinfo {author} {\bibfnamefont {S.-J.}\ \bibnamefont
  {Choi}}, \bibinfo {author} {\bibfnamefont {S.}~\bibnamefont {Park}},\ and\
  \bibinfo {author} {\bibfnamefont {H.-S.}\ \bibnamefont {Sim}},\ }\bibfield
  {title} {\bibinfo {title} {Geometric phase at a graphene edge: Scattering
  phase shift of {Dirac} fermions},\ }\href
  {https://doi.org/10.1103/PhysRevB.89.155412} {\bibfield  {journal} {\bibinfo
  {journal} {Phys. Rev. B}\ }\textbf {\bibinfo {volume} {89}},\ \bibinfo
  {pages} {155412} (\bibinfo {year} {2014})}\BibitemShut {NoStop}%
\bibitem [{\citenamefont {Xiao}\ \emph {et~al.}(2010)\citenamefont {Xiao},
  \citenamefont {Chang},\ and\ \citenamefont {Niu}}]{RevModPhys.82.1959}%
  \BibitemOpen
  \bibfield  {author} {\bibinfo {author} {\bibfnamefont {D.}~\bibnamefont
  {Xiao}}, \bibinfo {author} {\bibfnamefont {M.-C.}\ \bibnamefont {Chang}},\
  and\ \bibinfo {author} {\bibfnamefont {Q.}~\bibnamefont {Niu}},\ }\bibfield
  {title} {\bibinfo {title} {Berry phase effects on electronic properties},\
  }\href {https://doi.org/10.1103/RevModPhys.82.1959} {\bibfield  {journal}
  {\bibinfo  {journal} {Rev. Mod. Phys.}\ }\textbf {\bibinfo {volume} {82}},\
  \bibinfo {pages} {1959} (\bibinfo {year} {2010})}\BibitemShut {NoStop}%
\bibitem [{\citenamefont {Park}\ and\ \citenamefont
  {Marzari}(2011)}]{PhysRevB.84.205440}%
  \BibitemOpen
  \bibfield  {author} {\bibinfo {author} {\bibfnamefont {C.-H.}\ \bibnamefont
  {Park}}\ and\ \bibinfo {author} {\bibfnamefont {N.}~\bibnamefont {Marzari}},\
  }\bibfield  {title} {\bibinfo {title} {Berry phase and pseudospin winding
  number in bilayer graphene},\ }\href
  {https://doi.org/10.1103/PhysRevB.84.205440} {\bibfield  {journal} {\bibinfo
  {journal} {Phys. Rev. B}\ }\textbf {\bibinfo {volume} {84}},\ \bibinfo
  {pages} {205440} (\bibinfo {year} {2011})}\BibitemShut {NoStop}%
\bibitem [{\citenamefont {Zeng}\ and\ \citenamefont
  {Shen}(2022)}]{PhysRevB.106.094503}%
  \BibitemOpen
  \bibfield  {author} {\bibinfo {author} {\bibfnamefont {W.}~\bibnamefont
  {Zeng}}\ and\ \bibinfo {author} {\bibfnamefont {R.}~\bibnamefont {Shen}},\
  }\bibfield  {title} {\bibinfo {title} {Pure crossed {Andreev} reflection
  assisted transverse valley currents in $\alpha-\mathcal{T}_{3}$ lattices},\
  }\href {https://doi.org/10.1103/PhysRevB.106.094503} {\bibfield  {journal}
  {\bibinfo  {journal} {Phys. Rev. B}\ }\textbf {\bibinfo {volume} {106}},\
  \bibinfo {pages} {094503} (\bibinfo {year} {2022})}\BibitemShut {NoStop}%
\bibitem [{\citenamefont {Kato}\ \emph {et~al.}(2004)\citenamefont {Kato},
  \citenamefont {Myers}, \citenamefont {Gossard},\ and\ \citenamefont
  {Awschalom}}]{doi:10.1126/science.1105514}%
  \BibitemOpen
  \bibfield  {author} {\bibinfo {author} {\bibfnamefont {Y.~K.}\ \bibnamefont
  {Kato}}, \bibinfo {author} {\bibfnamefont {R.~C.}\ \bibnamefont {Myers}},
  \bibinfo {author} {\bibfnamefont {A.~C.}\ \bibnamefont {Gossard}},\ and\
  \bibinfo {author} {\bibfnamefont {D.~D.}\ \bibnamefont {Awschalom}},\
  }\bibfield  {title} {\bibinfo {title} {Observation of the spin {Hall} effect
  in semiconductors},\ }\href {https://doi.org/10.1126/science.1105514}
  {\bibfield  {journal} {\bibinfo  {journal} {Science}\ }\textbf {\bibinfo
  {volume} {306}},\ \bibinfo {pages} {1910} (\bibinfo {year}
  {2004})}\BibitemShut {NoStop}%
\bibitem [{\citenamefont {Ando}\ and\ \citenamefont
  {Saitoh}(2012)}]{ando2012observation}%
  \BibitemOpen
  \bibfield  {author} {\bibinfo {author} {\bibfnamefont {K.}~\bibnamefont
  {Ando}}\ and\ \bibinfo {author} {\bibfnamefont {E.}~\bibnamefont {Saitoh}},\
  }\bibfield  {title} {\bibinfo {title} {Observation of the inverse spin {Hall}
  effect in silicon},\ }\href {https://doi.org/10.1038/ncomms1640} {\bibfield
  {journal} {\bibinfo  {journal} {Nature communications}\ }\textbf {\bibinfo
  {volume} {3}},\ \bibinfo {pages} {629} (\bibinfo {year} {2012})}\BibitemShut
  {NoStop}%
\bibitem [{\citenamefont {Balakrishnan}\ \emph {et~al.}(2014)\citenamefont
  {Balakrishnan}, \citenamefont {Koon}, \citenamefont {Avsar}, \citenamefont
  {Ho}, \citenamefont {Lee}, \citenamefont {Jaiswal}, \citenamefont {Baeck},
  \citenamefont {Ahn}, \citenamefont {Ferreira}, \citenamefont {Cazalilla}
  \emph {et~al.}}]{balakrishnan2014giant}%
  \BibitemOpen
  \bibfield  {author} {\bibinfo {author} {\bibfnamefont {J.}~\bibnamefont
  {Balakrishnan}}, \bibinfo {author} {\bibfnamefont {G.~K.~W.}\ \bibnamefont
  {Koon}}, \bibinfo {author} {\bibfnamefont {A.}~\bibnamefont {Avsar}},
  \bibinfo {author} {\bibfnamefont {Y.}~\bibnamefont {Ho}}, \bibinfo {author}
  {\bibfnamefont {J.~H.}\ \bibnamefont {Lee}}, \bibinfo {author} {\bibfnamefont
  {M.}~\bibnamefont {Jaiswal}}, \bibinfo {author} {\bibfnamefont {S.-J.}\
  \bibnamefont {Baeck}}, \bibinfo {author} {\bibfnamefont {J.-H.}\ \bibnamefont
  {Ahn}}, \bibinfo {author} {\bibfnamefont {A.}~\bibnamefont {Ferreira}},
  \bibinfo {author} {\bibfnamefont {M.~A.}\ \bibnamefont {Cazalilla}}, \emph
  {et~al.},\ }\bibfield  {title} {\bibinfo {title} {Giant spin {Hall} effect in
  graphene grown by chemical vapour deposition},\ }\href
  {https://doi.org/10.1038/ncomms5748} {\bibfield  {journal} {\bibinfo
  {journal} {Nature communications}\ }\textbf {\bibinfo {volume} {5}},\
  \bibinfo {pages} {4748} (\bibinfo {year} {2014})}\BibitemShut {NoStop}%
\bibitem [{\citenamefont {Gao}\ \emph {et~al.}(2023)\citenamefont {Gao},
  \citenamefont {Cheng},\ and\ \citenamefont {Sun}}]{PhysRevB.108.024504}%
  \BibitemOpen
  \bibfield  {author} {\bibinfo {author} {\bibfnamefont {L.}~\bibnamefont
  {Gao}}, \bibinfo {author} {\bibfnamefont {Q.}~\bibnamefont {Cheng}},\ and\
  \bibinfo {author} {\bibfnamefont {Q.-F.}\ \bibnamefont {Sun}},\ }\bibfield
  {title} {\bibinfo {title} {Spin-valley dependent double {Andreev} reflections
  in the proximitized graphene/superconductor junction},\ }\href
  {https://doi.org/10.1103/PhysRevB.108.024504} {\bibfield  {journal} {\bibinfo
   {journal} {Phys. Rev. B}\ }\textbf {\bibinfo {volume} {108}},\ \bibinfo
  {pages} {024504} (\bibinfo {year} {2023})}\BibitemShut {NoStop}%
\bibitem [{\citenamefont {Akhmerov}\ and\ \citenamefont
  {Beenakker}(2008)}]{PhysRevB.77.085423}%
  \BibitemOpen
  \bibfield  {author} {\bibinfo {author} {\bibfnamefont {A.~R.}\ \bibnamefont
  {Akhmerov}}\ and\ \bibinfo {author} {\bibfnamefont {C.~W.~J.}\ \bibnamefont
  {Beenakker}},\ }\bibfield  {title} {\bibinfo {title} {Boundary conditions for
  {Dirac} fermions on a terminated honeycomb lattice},\ }\href
  {https://doi.org/10.1103/PhysRevB.77.085423} {\bibfield  {journal} {\bibinfo
  {journal} {Phys. Rev. B}\ }\textbf {\bibinfo {volume} {77}},\ \bibinfo
  {pages} {085423} (\bibinfo {year} {2008})}\BibitemShut {NoStop}%
\bibitem [{\citenamefont {Basko}(2009)}]{PhysRevB.79.205428}%
  \BibitemOpen
  \bibfield  {author} {\bibinfo {author} {\bibfnamefont {D.~M.}\ \bibnamefont
  {Basko}},\ }\bibfield  {title} {\bibinfo {title} {Boundary problems for
  {Dirac} electrons and edge-assisted {Raman} scattering in graphene},\ }\href
  {https://doi.org/10.1103/PhysRevB.79.205428} {\bibfield  {journal} {\bibinfo
  {journal} {Phys. Rev. B}\ }\textbf {\bibinfo {volume} {79}},\ \bibinfo
  {pages} {205428} (\bibinfo {year} {2009})}\BibitemShut {NoStop}%
\bibitem [{\citenamefont {Dyrdał}\ and\ \citenamefont
  {Barnaś}(2017)}]{Dyrdał_2017}%
  \BibitemOpen
  \bibfield  {author} {\bibinfo {author} {\bibfnamefont {A.}~\bibnamefont
  {Dyrdał}}\ and\ \bibinfo {author} {\bibfnamefont {J.}~\bibnamefont
  {Barnaś}},\ }\bibfield  {title} {\bibinfo {title} {Anomalous, spin, and
  valley {Hall} effects in graphene deposited on ferromagnetic substrates},\
  }\href {https://doi.org/10.1088/2053-1583/aa7bac} {\bibfield  {journal}
  {\bibinfo  {journal} {2D Materials}\ }\textbf {\bibinfo {volume} {4}},\
  \bibinfo {pages} {034003} (\bibinfo {year} {2017})}\BibitemShut {NoStop}%
\bibitem [{\citenamefont {Nagaosa}\ \emph {et~al.}(2010)\citenamefont
  {Nagaosa}, \citenamefont {Sinova}, \citenamefont {Onoda}, \citenamefont
  {MacDonald},\ and\ \citenamefont {Ong}}]{RevModPhys.82.1539}%
  \BibitemOpen
  \bibfield  {author} {\bibinfo {author} {\bibfnamefont {N.}~\bibnamefont
  {Nagaosa}}, \bibinfo {author} {\bibfnamefont {J.}~\bibnamefont {Sinova}},
  \bibinfo {author} {\bibfnamefont {S.}~\bibnamefont {Onoda}}, \bibinfo
  {author} {\bibfnamefont {A.~H.}\ \bibnamefont {MacDonald}},\ and\ \bibinfo
  {author} {\bibfnamefont {N.~P.}\ \bibnamefont {Ong}},\ }\bibfield  {title}
  {\bibinfo {title} {Anomalous {Hall} effect},\ }\href
  {https://doi.org/10.1103/RevModPhys.82.1539} {\bibfield  {journal} {\bibinfo
  {journal} {Rev. Mod. Phys.}\ }\textbf {\bibinfo {volume} {82}},\ \bibinfo
  {pages} {1539} (\bibinfo {year} {2010})}\BibitemShut {NoStop}%
\end{thebibliography}
\end{document}